\documentclass[aps,prd,groupedaddress,preprint,eqsecnum,nofootinbib]{revtex4}
\usepackage{graphicx,epsf,amssymb,amsbsy,amsfonts,amssymb,amsmath,physics,courier}

\def\be{\begin{equation}}
\def\ee{\end{equation}}

\def\bg{\bar{g}}
\def\beq{\begin{eqnarray}}\def\eeq{\end{eqnarray}}
\def\ba#1\ea{\begin{align}#1\end{align}}
\def\bg#1\eg{\begin{gather}#1\end{gather}}
\def\bm#1\em{\begin{multline}#1\end{multline}}
\def\bmd#1\emd{\begin{multlined}#1\end{multlined}}

\def\({\left(}
\def\){\right)}
\def\[{\left[}
\def\]{\right]}

\pdfoutput=1

\usepackage{tgbonum}

\usepackage{beton}

\usepackage[T1]{fontenc}

\begin{document}
\hfuzz 9pt
%\maketitle
\title{Null Infinity and Unitary Representation of The Poincare Group}
\author{Shamik Banerjee}
%\date{}                                           % Activate to display a given date or no date
\affiliation{Institute of Physics, \\ Sachivalaya Marg, Bhubaneshwar, India-751005 \\ and \\ Homi Bhabha National Institute, Anushakti Nagar, Mumbai, India-400085}

\email{banerjeeshamik.phy@gmail.com}
\begin{abstract}
Following Pasterski-Shao-Strominger we construct a new basis of states in the single-particle Hilbert space of massless particles as a linear combination of standard Wigner states. Under Lorentz transformation the new basis states transform in the Unitary Principal Continuous Series representation. These states are obtained if we consider the little group of a null momentum \textit{direction} rather than a null momentum. The definition of the states in terms of the Wigner states makes it easier to study the action of space-time translation in this basis. We show by taking into account the effect of space-time translation that \textit{the dynamics of massless particles described by these states takes place completely on the null-infinity of the Minkowski space}. We then second quantize the theory in this basis and obtain a manifestly Poincare invariant (field) theory of free massless particles living on null-infinity. This theory has unitary time evolution. The null-infinity arises in this case purely group-theoretically without any reference to bulk space-time. Action of BMS is particularly natural in this picture.

As a by-product we generalize the conformal primary wave-functions for massless particles in a way which makes the action of space-time translation simple. Using these wave-functions we write down a modified Mellin(-Fourier) transformation of the S-matrix elements. The resulting amplitude is Poincare covariant. Under Poincare transformation it transforms like products of primaries of inhomogeneous $SL(2,\mathbb{C})$ ($ISL(2,\mathbb{C})$) inserted at various points of null-infinity. $ISL(2,\mathbb{C})$ primaries are defined in the paper. 

\end{abstract}

%\preprint{}
\maketitle
\tableofcontents

\section{Introduction And Summary Of Main Results}

In $(3+1)$ dimensions the Lorentz group $SL(2,\mathbb{C})$ acts as the group of global conformal transformations on the celestial sphere ($S^2$) at null-infinity. A series of very interesting recent papers \cite{Pasterski:2016qvg,Pasterski:2017kqt,Pasterski:2017ylz} have constructed external state wave functions for massive and massless particles which transform like 2-D conformal-primary operators under $SL(2,\mathbb{C})$ (Lorentz) transformation. In particular, the detailed analysis of \cite{Pasterski:2017kqt} shows that the "Unitary principal continuous series representation" of the Lorentz group plays a central role in this whole construction. One of the interesting points about this basis of "conformal primary wave-functions" is that the 4-D scattering amplitudes, expressed in this basis, transform like the correlation function of products of $SL(2,\mathbb{C})$ primaries inserted at various points of the celestial sphere at null-infinity. This suggests a relation between 4-D scattering amplitudes and a 2-D Euclidean CFT defined on the celestial sphere. This 4D-2D correspondence is a hint of flat-space holography \cite{Pasterski:2016qvg,Pasterski:2017kqt,Pasterski:2017ylz,Cheung:2016iub,deBoer:2003vf, Bagchi:2016bcd,Bagchi:2016geg,Cardona:2017keg,Banerjee:2017jeg,Schreiber:2017jsr}.

In this paper we further explore this construction by looking at the single-particle Hilbert space of \textit{massless} particles. Following \cite{Pasterski:2016qvg,Pasterski:2017kqt,Pasterski:2017ylz} we define a new basis in the single particle Hilbert space which is related to the standard Wigner-states by Mellin transformation. These new states transform in the "Unitary principal continuous series" representation of $SL(2,\mathbb{C})$ (Lorentz group). Geometrically speaking these new states are obtained if, instead of a null momentum vector, we consider the \textit{little-group of a null momentum direction}. A null momentum direction is defined as the \textit{set of all null momentum vectors which point in the same direction} and so can be represented by a \textit{single point of the celestial sphere in the momentum space}. These states are naturally written as $\ket{h,\bar h, z,\bar z}$ where $(z,\bar z)$ are the stereographic coordinates of the celestial sphere in the momentum space. $(h,\bar h)$ are little-group indices defined as,
\be\nonumber
h = \frac{1+i\lambda - \sigma}{2} , \ \bar h = \frac{1+i\lambda + \sigma}{2} 
\ee
where $\lambda\in\mathbb{R}$ and $\sigma$ is the helicity of the massless particle. It is shown in the paper that under a unitary $SL(2,\mathbb{C})$ (Lorentz) transformation $U(\Lambda)$ the state transforms as,
\be\nonumber
U(\Lambda) \ket{h, \bar h, z, \bar z} = \frac{1}{(cz+d)^{2h}} \frac{1}{(\bar c \bar z + \bar d)^{2 \bar h}} \ket{h \ , \bar h \ ,\frac{az+b}{cz+d} \ , \frac{\bar a \bar z + \bar b}{\bar c \bar z + \bar d}}, \ 
\Lambda =
\begin{pmatrix}
a & b \\
c & d
\end{pmatrix} \in{SL(2,\mathbb{C})},
\ee

These states are somewhat analogous to position eigenkets in non-relativistic quantum mechanics. In the state $\ket{h,\bar h,z,\bar z}$ the massless particle can be thought of as sitting at the point $(z,\bar z)$ of the celestial sphere in the momentum space and $(h,\bar h)$ are internal quantum numbers. These states form a complete set.

Now the definition of the states $\ket{h,\bar h,z,\bar z}$ in terms of the Wigner-states makes it easier to study the action of \textit{space-time translation} on them. It turns out that the simplest way to represent the action of the space-time translation on these states is to first consider the "position eigenket in Heisenberg-picture" defined as, 
\be\nonumber
\ket{h,\bar h,u,z,\bar z} = e^{iHu} \ket{h,\bar h, z, \bar z}
\ee
where $H$ is the Hamiltonian. These states are analogous to the states $\ket{t,\vec x}$ in non-relativistic quantum mechanics. The action of the \textit{Inhomogeneous $SL(2,\mathbb{C})$} ($ISL(2,\mathbb{C})$) or the Poincare group on the Heisenberg-picture states are shown to be,
\begin {equation}\nonumber
U(\Lambda) \ket{h, \bar h, u, z, \bar z} = \frac{1}{(cz+d)^{2h}} \frac{1}{(\bar c \bar z + \bar d)^{2 \bar h}} \ket{h \ , \bar h \ , \frac{u \ (1+ z \bar z)}{|az+b|^2 + |cz+d|^2}, \frac{az+b}{cz+d} \ , \frac{\bar a \bar z + \bar b}{\bar c \bar z + \bar d}}
\end{equation}
\begin{equation}\nonumber
e^{-il.P} \ket{h, \bar h, u, z, \bar z} = \ket{h, \bar h, u+ f(z, \bar z, l), z, \bar z}
\end{equation}
where $l^{\mu}$ is the space-time translation vector and 
\begin{equation}\nonumber
f(z, \bar z, l) = \frac{(l^0 - l^3) - (l^1 - i l^2) z - (l^1 + i l^2) \bar z + (l^0 + l^3) z \bar z}{1 + z \bar z}
\end{equation}
So a massless particle described by the quantum state $\ket{h,\bar h, z,\bar z}$ can be thought of as living in a space(-time) with three coordinates given by $(u,z,\bar z)$. The action of the Poincare group on the $(u,z,\bar z)$ space(-time) is given by,
\begin{equation}\nonumber
\Lambda (u,z, \bar z) =  \bigg( \frac{u \ (1+ z \bar z)}{|az+b|^2 + |cz+d|^2} \ , \frac{az+b}{cz+d} \ , \frac{\bar a \bar z + \bar b}{\bar c \bar z + \bar d}\bigg)
\end{equation}
\begin{equation}\nonumber
T(l) (u, z, \bar z) = (u + f(z, \bar z, l), z, \bar z)
\end{equation}
Here \textit{the coordinate $u$ is time-like} because we get it from Hamiltonian evolution of the state.  

Now, \textit{as far as the action of the Poincare group is concerned}, we can identify the three dimensional space-time as the null-infinity in Minkowski space with $(u,z,\bar z)$ identified as the Bondi coordinates at infinity. So from now on we will refer to this space as null-infinity but there is of course no "bulk space-time" metric here. The null-infinity arises purely group-theoretically and the dynamics of a free massless particle takes place in this space.

We then second-quantize the theory using the basis states $\ket{h,\bar h,z,\bar z}$ and write down \textit{Heisenberg-Picture creation and annihilation fields which live on the null-infinity}. So a manifestly Poincare-invariant and unitary many-body theory of free massless particles can be formulated completely on ($(u,z,\bar z)$ space-time) null-infinity \textit{without any reference to bulk space-time}. 

The transition from the momentum space to null-infinity is \textit{direct}. $(z,\bar z)$ start as coordinates in the momentum-space but once we take into account the dynamics, $(z,\bar z)$ together with the time coordinate $u$ transmute into space-time coordinates, albeit of the boundary. This is also consistent with the current understanding of the relation between infrared structure of gauge and gravity theories and asymptotic symmetries of flat space-time \cite{Strominger:2013jfa,Strominger:2013lka,He,Kapec:2014opa, Strominger:2014pwa,Kapec:2016jld,Campiglia:2014yka,Campiglia:2015yka,Campiglia:2015kxa,Kapec:2017gsg,Strominger:2017zoo,Weinberg:1965nx,Cachazo:2014fwa,Sen:2017xjn,Sen:2017nim,Laddha:2017ygw,Chakrabarti:2017ltl,Chakrabarti:2017zmh,Laddha:2017vfh,Hawking:2016msc}.

This free theory has some interesting properties. For example, the transition amplitude given by, 
\begin{equation}\nonumber
\begin{aligned}
\bra{h,\bar h, z,\bar z} e^{-iH(u-u')} \ket{h',\bar h',z',\bar z'} 
&= \bra{h,\bar h,u,z,\bar z}\ket{h',\bar h',u',z',\bar z'} \\
&= \frac{\delta_{\sigma \sigma'}}{2\pi} \frac{\Gamma\big(i(\lambda' - \lambda)\big)}{(1 + z \bar z)^{i(\lambda' - \lambda)}} \frac{\delta^2(z' -z)}{\big(-i(u'-u + i0+)\big)^{i(\lambda' - \lambda)}}
\end{aligned}
\end{equation}

is $SL(2,\mathbb{C})$ (Lorentz) covariant and is manifestly invariant under space-time translation because of the Dirac delta function. In fact the transition amplitude retains its form under more general "BMS supertranslation" given by $(u,z,\bar z)\rightarrow (u + g(z,\bar z),z,\bar z)$ where $g(z,\bar z)$ is an arbitrary smooth function on the sphere. So we get a hint of supertranslation in a \textit{purely non-gravitational} context. We discuss these things in detail in the paper.

This paper consists of two parts. In the first part we do the $(2+1)$ case as a warm-up and also to check the correctness the procedure. In the first part we do not discuss the action of space-time translation. The second part consists of the $(3+1)$ dimensional case which is our main interest. The two parts are more or less independent. \\

\textit{Throughout this paper we will use the mostly positive metric signature and use the conventions of Weinberg \cite{Weinberg:1995mt}}.

\section{Lorentz Group In (2+1) Dimensions}
In $(2+1)$ dimensions Lorentz group is $SO(2,1)$. In this paper we shall consider $SL(2,\mathbb{R})$ which is the double cover of $SO(2,1)$. $SL(2,\mathbb{R})$ is the group of real two by two matrices with determinant $1$. The group elements will be denoted by,

\begin{equation}
\Lambda =
\begin{pmatrix}
a & b \\
c & d
\end{pmatrix},  \
ad-bc=1, \  (a,b,c,d)\in\mathbb{R}
\end{equation}

Let us now consider a momentum vector $P = (P^0,P^1,P^2)$ and associate a real symmetric two by two matrix, denoted by the same letter $P$, defined as

\begin{equation}
P = 
\begin{pmatrix}
P^0 - P^2 & P^1 \\
P^1 & P^0 + P^2
\end{pmatrix}, \
\det P = (P^0)^2 - (P^1)^2 - (P^2)^2 = -P^2
\end{equation}

So the determinant gives the norm of the three vector $P$. \\

The $SL(2,\mathbb{R})$ matrix $\Lambda$ acts on $P$ according to the following rule,

\begin{equation}
P' =  \Lambda P \Lambda^T
\end{equation}

Under this transformation $P'$ is again a real symmetric matrix and, $\det P =  \det P'$ . Therefore $\{\pm\Lambda\}$ induces a Lorentz transformation of the three vector $P$. 

\subsection{Null Momenta}
Let us now specialize to the case of null momenta. When $P$ is null we can define a real variable $z$ as, 

\begin{equation}
z=\frac{P^1}{P^0 + P^2}
\end{equation}

The variable $z$ has the property that when $P$ transforms under Lorentz transformation as $P\rightarrow P' = \Lambda P \Lambda^T$ , $z$ transforms as,

\begin{equation}\label{LT3}
z \rightarrow z' = \Lambda z = \frac{P^{1'}}{P^{0'} + P^{2'}} = \frac{az+b}{cz+d}
\end{equation}

The geometrical interpretation of the variable $z$ is that it parametrizes the \textit{space of null momentum directions} rather than a specific null vector. To see this let us consider the set of all null vectors which point in the positive $2$-direction. These vectors have the form  $k^{\mu}=(E,0,E)$ where $E>0$. It is easy to check that the set of all such null vectors get mapped to the point $z=0$. 

Now let us make a rotation in the $1-2$ plane through an angle $\theta$. Under this transformation $k^{\mu}\rightarrow k'^{\mu} = (E, E \sin\theta, E \cos\theta)$. This is the most general form of a null vector which makes an angle $\theta$ with the positive $2$-axis. As a result of this rotation, 

\begin{equation}\label{utheta}
z=0 \rightarrow z' = \tan\frac{\theta}{2}
\end{equation}

So we can see that the parameter $z$ depends only on the direction of the null ray and the set of all null vectors that point in the same direction correspond to a \textit{single} value of $z$. In fact, once we choose the set $\{k^{\mu}=(E,0,E) , E>0\}$ as the \textit{reference null direction with $(z=0, \theta=0)$} then there is an invertible correspondence between the space of null directions and values of $z$ given by Eq-($\ref{utheta}$). 

Geometrically the space of null momentum directions in (2+1) dimensions is a circle parametrized by the angle $\theta$ and $z$ is the stereographic coordinate on the circle. The Lorentz group $SL(2,\mathbb{R})$ acts on the $z$ coordinate according to Eq-$\ref{LT3}$.

\subsection{Standard Rotation}

A standard rotation, $R(z)$, is a rotation in the $1-2$ plane which takes the \textit{reference null direction}, $\bigg\{k^{\mu}=(E,0,E), E>0 \big | z=0\bigg\}$ to the null direction $\bigg\{ k^{\mu} = (E,E\sin\theta,E\cos\theta), E>0 \big | z= \tan\frac{\theta}{2}\bigg\}$. In matrix form, 

\begin{equation}
R(z) = \pm
\begin{pmatrix}
\cos\frac{\theta}{2} & \sin\frac{\theta}{2} \\\\
-\sin\frac{\theta}{2} & \cos\frac{\theta}{2}
\end{pmatrix} , \
z= \tan\frac{\theta}{2}
\end{equation}

\subsection{Little Group of A Null Momentum \textit{Direction}}

In the standard Wigner construction \textit{little group} of a \textit{null vector} plays a central role. We want to consider instead the \textit{little group of a null direction}. This is the subgroup of the Lorentz group which does not change the \textit{direction} of a null vector. As a result, \textit{the set of all null vectors which point in the same direction, is closed under the action of this subgroup}. Let us now explicitly write down the subgroup in (2+1) dimensions. 

The generators of $SL(2,\mathbb{R})$ can be written as $(J_0,K_1,K_2)$, where $J_0$ is the generator of rotation in the $1-2$ plane, $K_1$ is the generator of boost in the $1$-direction and $K_2$ is the generator of boost in the $2$-direction. Algebra satisfied by these generators is, 

\begin{equation}
[J_0 , K_1] = iK_2 , \ [J_0,K_2] = -iK_1, \ [K_1,K_2] = -i J_0
\end{equation} 

Let us now consider the little group of the \textit{reference null direction} given by the set $\bigg\{k^{\mu}=(E,0,E), E>0 \big | z=0\bigg\}$. The little group of the \textit{null vector}, $(1,0,1)$, is generated by the element, $A=-J_0-K_1$. It is well known that the one-parameter subgroup generated by $A$ is non-compact and isomorphic to the group of translation. Now in order to get the little group of the reference null direction we just have to add the boost generator $K_2$ to this. Therefore, \textit{the little group, $D(z=0)$, of the reference null direction, $\bigg\{k^{\mu}=(E,0,E), E>0 \big | z=0\bigg\}$, is generated by two elements, $A = -J_0-K_1$ and $K_2$}. One can check that, $[K_2,A] = iA$. Similarly the little group of the null direction, $\bigg\{ k^{\mu} = (E,E\sin\theta,E\cos\theta), E>0 \big | z= \tan\frac{\theta}{2}\bigg\}$, is given by conjugation, $ D(z) = R(z)D(z=0)R(z)^{-1}$.

\subsection{Hilbert Space Representation}

Let us consider massless single particle quantum states in $(2+1)$ dimensions. The single particle states are denoted by $\ket{p}$ with $p^2=0$. The Lorentz-invariant inner product between the states is given by, $\bra{p_1}\ket{p_2} = (2\pi)^2 \ 2|\vec p_1 | \ \delta^2(\vec p_1 - \vec p_2)$. \\

Let $\Lambda$ be an arbitrary Lorentz transformation and $U(\Lambda)$ its unitary representation in the Hilbert space which acts on $\ket{p}$ as, 

\begin{equation}\label{LT}
U(\Lambda) \ket{p}  = \ket{\Lambda p}
\end{equation}

If we take $\Lambda$ to be the boost in the $2$-direction with velocity $v$ then
\begin{equation}
U(\Lambda) \ket{E,0,E} = \ket{e^{-\eta}E,0,e^{-\eta}E} 
\end{equation}
where $\eta$ is the rapidity defined as, $\tanh\eta = v$. 

Now the $SL(2,\mathbb{R})$ matrix which generates boost in the 2-direction is given by , 
\begin{equation}
B_2(\eta) = \pm
\begin{pmatrix}
e^{\frac{\eta}{2}} & 0 \\\\
0 & e^{-\frac{\eta}{2}}
\end{pmatrix}
\end{equation}

The transformation of the $z$ variable under boost in the $2$-direction is, $z \rightarrow e^{\eta} z$. So it is a scale transformation for which $z=0$ is a fixed point. This is another way of seeing that $K_2$ is a generator of the little group of the reference null direction $\bigg\{k^{\mu}=(E,0,E), E>0 \big | z=0\bigg\}$. 

Let us mention one more important point. Here we are considering single particle states $\ket{p}$ annihilated by the single non-compact generator $A = - J_0 - K_1$ of the \textit{little group of a null vector}, i.e,
\begin{equation}
A \ket{E,0,E} =0, E>0
\end{equation}

%Since $[K_2,A]=iA$, one can easily check that, $A\ket{E,0,E} =0$ for all $E>0$. \\

\subsubsection{Change of Basis}
The works of \cite{Pasterski:2017kqt, Pasterski:2016qvg, Cheung:2016iub,deBoer:2003vf,Pasterski:2017ylz} suggest the introduction of the following states parametrized by a complex number $\Delta$,
\begin{equation}
\ket{\Delta , z=0}  \ = \frac{1}{(2\pi)^{\frac{3}{2}}} \int_0^{\infty} dE \ E^{\Delta -1} \ket{E,0,E}
\end{equation}

Now if we apply a boost in the $2$-direction with rapidity $\eta$ we get,
\begin{equation}\label{beigenvalue}
U\big(B_2(\eta)\big)\ket{\Delta, z=0} =\frac{1}{(2\pi)^{\frac{3}{2}}} \int_0^{\infty} dE \ E^{\Delta -1} \ket{e^{-\eta}E,0,e^{-\eta}E} \ = e^{\eta\Delta} \ket{\Delta, z=0}
\end{equation}

Let us now define states $\ket{\Delta,z}$ which are related to $\ket{\Delta, z=0}$ by standard rotation $R(z)$ in the $1-2$ plane, i.e, 

\begin{equation}\label{define}
\boxed{\ket{\Delta, z}:= \frac{1}{(1+z^2)^{\Delta}} \ U\big(R(z)\big) \ket{\Delta, z=0} = N_{\Delta}(z) U(R(z))\ket{\Delta, z=0}}
\end{equation} 

We can write down the states explicitly as, 
\begin{equation}
\ket{\Delta,z} = \frac{1}{(2\pi)^{\frac{3}{2}}}\frac{1}{(1+z^2)^{\Delta}} \int_{0}^{\infty} dE E^{\Delta-1} \ket{E, E\sin\theta,E \cos\theta}
\end{equation}

The inner product between two such states can be written as, 
\begin{equation}
\begin{aligned}
\bra{\Delta_2,z_2}\ket{\Delta_1,z_1}  \\
&= \frac{1}{(2\pi)^3} \frac{1}{(1+z_2^2)^{\Delta_2^*} (1+ z_1^2)^{\Delta_1}} \times \\
& \int_{0}^{\infty}\int_{0}^{\infty} dE_2 dE_1 {E_2}^{\Delta_2^{*} -1} {E_1}^{\Delta_1 -1} \bra{E_2,E_2\sin\theta_2,E_2\cos\theta_2}\ket{E_1,E_1\sin\theta_1,E_1\cos\theta_1}
\end{aligned}
\end{equation}

where $*$ denotes complex conjugation. This integral is convergent if we choose $\Delta$ to be of the form $(\frac{1}{2} + i\lambda , \ \lambda\in\mathbb{R})$ \cite{Pasterski:2017kqt}. With this choice of $\Delta$ the states that we have defined are delta function normalizable, i.e, 

\begin{equation}
\bra{\Delta_2, z_2}\ket{\Delta_1, z_1} = \delta(\lambda_2 - \lambda_1) \delta(z_2 - z_1)
\end{equation}

where $\Delta_i = \frac{1}{2} + i\lambda_i$.

\subsubsection{Action of $U(\Lambda)$}

Let us now calculate $U(\Lambda)\ket{\Delta,z}$ where $\Lambda$ is an arbitrary Lorentz transformation. If we use the definition of the state given in Eq-(\ref{define}) we get,  
\begin{equation}\label{L1}
\begin{aligned}
U(\Lambda)\ket{\Delta,z} 
& = N_{\Delta}(z) U(\Lambda) \ U(R(z)) \ket{\Delta, z=0}  \\
&  = N_{\Delta}(z) U(R(\Lambda z)) \ U^{-1}(R(\Lambda z)) \ U(\Lambda) \ U(R(z)) \ket{\Delta, z=0} \\
& = N_{\Delta}(z) U(R(\Lambda z)) \ W(\Lambda,z) \ket{\Delta, z=0}
\end{aligned}
\end{equation}

%where $N_{\Delta}(u) = \frac{1}{(2\pi)^{\frac{3}{2}}}\frac{1}{(1+u^2)^{\Delta}}$, is the normalization factor. \\

where $W(\Lambda,z)$ is the Lorentz transformation given by, 
\begin{equation}\label{L2}
W(\Lambda, z) = U^{-1}(R(\Lambda z)) \ U(\Lambda) \ U(R(z)) = U(R^{-1}(\Lambda z) \Lambda  R(z))
\end{equation}

Now using the definition of the state $\ket{\Delta, z=0}$ we get, 
\begin{equation}\label{LG}
W(\Lambda, z) \ket{\Delta, z=0} = \frac{1}{(2\pi)^{\frac{3}{2}}} \int_0^{\infty} dE \ E^{\Delta -1} W(\Lambda,z) \ket{E,0,E}
\end{equation}

So we have to find out the effect of the transformation $R^{-1}(\Lambda z) \Lambda R(z)$ on the set of null vectors of the form $(E,0,E)$, pointing in the positive $2$-direction which is our reference direction. Since Lorentz transformation is linear it is sufficient to consider its effect on the vector $k^{\mu} = (1,0,1)$. The transformations act on $k^{\mu}$ in the following way :

1) $R(z)$ sends the vector $k^{\mu} = (1,0,1)$ to the vector $k'^{\mu} = (1, \sin\theta, \cos\theta)$. The corresponding $z$ transforms from $z=0$ to $z= \tan\frac{\theta}{2}$. 

2) Now we make an arbitrary Lorentz transformation $\Lambda$ given by the $SL(2,\mathbb{R})$ matrix, 

\begin{equation}
\Lambda = 
\begin{pmatrix}
a & b \\
c & d
\end{pmatrix} , \
ad-bc = 1
\end{equation}

This transformation sends $k'$ to $\Lambda k'$ and $z = \tan\frac{\theta}{2}$ to 

\begin{equation}\label{tangle}
\Lambda z = \frac{a\tan\frac{\theta}{2} +b}{c\tan\frac{\theta}{2} + d} = \tan\frac{\Lambda\theta}{2}
\end{equation}

where $\Lambda\theta$ is the angle made by $\Lambda k'$ with the positive $2$-direction. 

The components of $\Lambda k'$ are given by, 

\begin{equation}
(\Lambda k')^0 = \frac{(cz+d)^2 + (az+b)^2}{1+z^2} , \  z= \tan\frac{\theta}{2}
\end{equation}

\begin{equation}
(\Lambda k')^1 = \frac{2(az+b)(cz+d)}{1+z^2} , \  z= \tan\frac{\theta}{2}
\end{equation}

\begin{equation}
(\Lambda k')^2 = \frac{(cz+d)^2 - (az+b)^2}{1+z^2} , \  z= \tan\frac{\theta}{2}
\end{equation}

We also have the relation,
\begin{equation}
(\Lambda k')^2 = (\Lambda k')^0 \cos\Lambda\theta , \ (\Lambda k')^1 = (\Lambda k')^0 \sin\Lambda\theta
\end{equation}

%Eq-($\ref{tangle}$) follows from the expression of the components of the vector $\Lambda k'$, given by, 

%then we get back Eq-($\ref{tangle}$). \\

3) Now the final rotation $R^{-1}(\Lambda z)$ rotates the vector $\Lambda k'$ in 1-2 plane through an angle $-\Lambda\theta$ and the resulting vector points again in the positive 2-direction. Similarly the rotation brings $\Lambda z = \tan\frac{\Lambda\theta}{2}$ back to $z=0$ which is the reference null direction $\bigg\{ (E,0,E) , E>0 \bigg | z=0 \bigg\}$. The form of the matrix $R^{-1}(\Lambda z)$ is ,

\begin{equation}
R^{-1}(\Lambda z) = \pm
\begin{pmatrix}
\cos\frac{\Lambda\theta}{2} & -\sin\frac{\Lambda\theta}{2} \\\\
\sin\frac{\Lambda\theta}{2} & \cos\frac{\Lambda\theta}{2}
\end{pmatrix} , \
\Lambda z= \tan\frac{\Lambda\theta}{2}
\end{equation} \\

Therefore the resulting vector can be written as,
\begin{equation}
\bigg(R^{-1}(\Lambda z) \Lambda R(z)\bigg) (1,0,1) = \bigg(\frac{(cz+d)^2 + (az+b)^2}{1+z^2}, \ 0 , \ \frac{(cz+d)^2 + (az+b)^2}{1+z^2}\bigg)
\end{equation}

Here we have made use of the facts that the $0$-th component of a vector does not change under rotation in the $1-2$ plane and the transformed vector is a null vector. 

For a general vector $k^{\mu}$ of the form $(E,0,E)$ we can write, 
\begin{equation}\label{boost}
\bigg(R^{-1}(\Lambda z) \Lambda R(z)\bigg) (E,0,E) = B_2(\eta)(E,0,E) = (e^{-\eta} E, 0 , e^{-\eta}E), \ E>0
\end{equation}

where we have defined the boost factor $e^{-\eta}$ as, 

\begin{equation}
\boxed{e^{-\eta} = \frac{(cz+d)^2 + (az+b)^2}{1+z^2} = (cz+d)^2 \ \frac{1+ (\Lambda z)^2}{1+z^2}, \  \Lambda z = \frac{az+b}{cz+d}}
\end{equation} 

With this information we can now derive the transformation law as,
\begin{equation}
\begin{aligned}
U(\Lambda)\ket{\Delta,z} 
& = N_{\Delta}(z) U(R(\Lambda z)) \ \underbrace{U^{-1}(R(\Lambda z)) \ U(\Lambda) \ U(R(z))} \ket{\Delta, z=0} \\
%& = N_{\Delta}(u) U(R(\Lambda u)) W(\Lambda,u) \ket{\Delta,u=0}  \\
& = N_{\Delta}(z) U(R(\Lambda z)) \ W(\Lambda, z) \ket{\Delta,z=0} \\
& = N_{\Delta}(z) U(R(\Lambda z)) \ e^{\eta\Delta} \ket{\Delta,z=0}, \ \text{We have used Eq-$\ref{LG}$, Eq-$\ref{LT}$, Eq-$\ref{beigenvalue}$} \\
& = e^{\eta\Delta} \frac{N_{\Delta}(z)}{N_{\Delta}(\Lambda z)} \ \underbrace{N_{\Delta}(\Lambda z) U(R(\Lambda z)) \ket{\Delta,z=0}} \\
& = e^{\eta\Delta} \frac{N_{\Delta}(z)}{N_{\Delta}(\Lambda z)} \ket{\Delta, \Lambda z} , \ \text{We have used Eq-$\ref{define}$} \\
& = \frac{1}{(cz+d)^{2\Delta}} \ \frac{(1+z^2)^{\Delta}}{(1+ (\Lambda z)^2)^{\Delta}} \ \frac{(1+(\Lambda z)^2)^{\Delta}}{(1+z^2)^{\Delta}} \ \ket{\Delta, \Lambda z} \\
& = \frac{1}{(cz+d)^{2\Delta}} \ket{\Delta, \Lambda z}
\end{aligned}
\end{equation} \\

So a general Lorentz transformation $\Lambda$ acts on the states $\ket{\Delta,z}$ like, 

\begin{equation}\label{multiplier}
\boxed{
U(\Lambda) \ket{\Delta, z} = \frac{1}{(cz+d)^{2\Delta}} \ket{\Delta, \Lambda z}, \ 
\Lambda =
\begin{pmatrix}
a & b \\
c & d 
\end{pmatrix} \in SL(2,\mathbb{R})}
\end{equation}

Now using this it is easy to see that the states $\big\{\ket{\Delta,\Lambda}\big\}$, for a fixed $\Delta$, form a representation of the Lorentz group. In order to see that let us consider two Lorentz transformations $\Lambda_1$ and $\Lambda_2$. 

\begin{equation}
\begin{aligned}
U(\Lambda_2) U(\Lambda_1) \ket{\Delta,z} 
& = U(\Lambda_2) \bigg(\frac{1}{(c_1 z + d_1)^{2\Delta}} \ket{\Delta, \Lambda_1 z} \bigg) \\
& = \frac{1}{(c_1 z + d_1)^{2\Delta}}  \ U(\Lambda_2) \ket{\Delta, \Lambda_1 z} \\
& = \frac{1}{(c_1 z + d_1)^{2\Delta}} \ \frac{1}{(c_2 (\Lambda_1 z) + d_2)^{2\Delta}} \ket{\Delta, \Lambda_2 \Lambda_1 z} \\
& = \frac{1}{(c_{\Lambda_2\Lambda_1} z  + d_{\Lambda_2\Lambda_1})^{2\Delta}} \ket{\Delta, \Lambda_2\Lambda_1 z} \\\\
& = U(\Lambda_2\Lambda_1) \ket{\Delta, z}
\end{aligned}
\end{equation}

Where, 

\begin{equation}
\Lambda_2\Lambda_1 =
\begin{pmatrix}
a_{\Lambda_2\Lambda_1} & b_{\Lambda_2\Lambda_1} \\
c_{\Lambda_2\Lambda_1} & d_{\Lambda_2\Lambda_1}
\end{pmatrix}
= 
\begin{pmatrix}
a_2 & b_2 \\
c_2 & d_2 
\end{pmatrix}
\
\begin{pmatrix}
a_1 & b_1 \\
c_1 & d_1
\end{pmatrix}
\end{equation}

Therefore, 
\begin{equation}
U(\Lambda_2)U(\Lambda_1) = U(\Lambda_2\Lambda_1)
\end{equation}

and so $U(\Lambda)$ form a representation of the Lorentz group when acting on the states $\big\{\ket{\Delta,z}\big\}$ according to Eq-$\ref{multiplier}$. So we have representations of the Lorentz group parametrized by \big\{$\Delta = \frac{1}{2}+i\lambda, \lambda\in\mathbb{R}\big\}$.

This is also a unitary representation as one can easily check that, 

\begin{equation}
\big(U(\Lambda)\ket{\Delta',z'} , U(\Lambda)\ket{\Delta,z}\big) = \big(\ket{\Delta',z'}, \ket{\Delta,z}\big)
\end{equation}

The representation labelled by a fixed $\Delta$ is known as the "Unitary principal continuous series" representation of the Lorentz group \cite{bargman} which has also appeared in several recent investigations of CFT \cite{Gadde:2017sjg,Hogervorst:2017sfd,Simmons-Duffin:2017nub}. 

\section{Representation on Wave Functions (Packets)}

The inner product between the states is given by,

\begin{equation}
\bra{\Delta',z'}\ket{\Delta,z} = \delta(\lambda' - \lambda) \delta(z'-z), \ \Delta = \frac{1}{2} + i\lambda, \ \lambda\in\mathbb{R}
\end{equation}

So we can write the completeness relation as,

\begin{equation}
\ket{\Psi} = \int_{-\infty}^{\infty} \int_{-\infty}^{\infty} d\lambda \ dz \ \ket{\Delta,z}\bra{\Delta,z}\ket{\Psi} =\int_{-\infty}^{\infty} \int_{-\infty}^{\infty} d\lambda \ dz \ \ket{\Delta,z} \Psi(\Delta,z) 
\end{equation}

where $\ket{\Psi}$ is an arbitrary massless one-particle state in the QFT. Using this the inner product between two states can be written as,

\begin{equation}\label{innerproduct}
\bra{\Phi}\ket{\Psi} =  \int_{-\infty}^{\infty} \int_{-\infty}^{\infty} d\lambda \ dz \ \Phi^{*}(\Delta,z) \ \Psi(\Delta,z)
\end{equation}

where $*$ denotes complex conjugation. We want to find out the effect of an arbitrary Lorentz transformation $\Lambda$ on the wave function $\Psi(\Delta, z)$. In order to do that we can write,

\begin{equation}
\begin{aligned}
\big(U(\Lambda^{-1})\Psi\big)(\Delta,z) 
& = \big(\ket{\Delta,z}, U(\Lambda^{-1})\ket{\Psi}\big) \\
& = \big(U(\Lambda^{-1})^{\dagger} \ket{\Delta,z} , \ket{\Psi}\big) = \big(U(\Lambda) \ket{\Delta,z} , \ket{\Psi}\big) \\
& = \big(U(\Lambda) \ket{\Delta,z} , \ket{\Psi}\big) \\
& = \bigg(\frac{1}{(cz + d)^{2\Delta}} \ket{\Delta, \Lambda z}, \ket{\Psi}\bigg) \\
& = \frac{1}{(cz + d)^{2\Delta^*}} \ \Psi(\Delta, \Lambda z) , \ \Delta^* = \frac{1}{2} - i\lambda
\end{aligned}
\end{equation}

Therefore the action of the Lorentz group on the wave-functions is, 

\begin{equation}\label{waverep}
\boxed{\big(U(\Lambda^{-1})\Psi\big)\big(\Delta,z\big) = \frac{1}{(cz + d)^{2\Delta^*}} \ \Psi\bigg(\Delta,\frac{az+b}{cz+d}\bigg) , \ \Delta^* = \frac{1}{2} - i\lambda}
\end{equation}

%The natural inner product for these projected wave functions is the one induced from Eq-$\ref{innerproduct}$ and is given by, 

%\begin{equation}\label{subinner}
%\big(\Phi_{\Delta}, \Psi_{\Delta}\big) = \int_{-\infty}^{\infty} du \ \overline{\Phi_{\Delta}(u)} \ \Psi_{\Delta}(u)
%\end{equation}

%We will use the transformation given in Eq-$\ref{waverep}$ to find out differential operator realization of Lorentz generators. This will also give us the value of the Casimir for the $\Delta$-representation. 

\subsection{Differential Operators and Casimir}

The Lorentz group $SL(2,\mathbb{R})$ is the group of conformal transformations of the line (or circle) parametrized by $z$. So let us consider dilatation, special conformal transformation and translation separately. 

\subsubsection{Translation} 

Let us consider translation given by,

\begin{equation}
T(\alpha) z = z' = z+\alpha, \ 
\Lambda =
\begin{pmatrix}
a & b \\
c & d
\end{pmatrix}
=
\begin{pmatrix}
1 & \alpha \\
0 & 1
\end{pmatrix}
\end{equation}

Now after a straightforward calculation, which we describe in the Appendix, we get,

\begin{equation}
U\big(T(\alpha)\big) = e^{i \alpha\big(J_0 - K_1\big)} = e^{\alpha L_{-1}}
\end{equation}

So,

\begin{equation}
(L_{-1})^{\dagger} = - L_{-1}
\end{equation}

Now using Eq-$\ref{waverep}$, we get,

\begin{equation}
L_{-1} = i (J_0 - K_1) = - \frac{d}{dz}
\end{equation}

%Now it is obvious from the inner product (Eq-$\ref{subinner}$) that $L_{-1}$ is antihermitian, as is required for the unitarity of the  representation of the Lorentz group. 

\subsubsection{Special Conformal Transformation (SCT)}

Let us consider the SCT given by,

\begin{equation}
S(\alpha) z = z' = \frac{z}{1- \alpha z}  \ ,  \
\Lambda =
\begin{pmatrix}
a & b \\
c & d
\end{pmatrix}
=
\begin{pmatrix}
1 & 0 \\
-\alpha & 1
\end{pmatrix}
\end{equation}

After a starightforward calculation we get,

\begin{equation}
U\big(S(\alpha)\big) = e^{i \alpha (J_0 + K_1)} = e^{\alpha L_1}
\end{equation}

So,

\begin{equation}
(L_1)^{\dagger} = - L_1
\end{equation}

In differential operator form we get,

\begin{equation}
L_1 = i (J_0 + K_1) = - z^2 \frac{d}{dz} - 2 \Delta^* z
\end{equation}

%Now it is easy to check using the inner product (Eq-$\ref{subinner}$) and the value of $\Delta$ given by, $\Delta = \frac{1}{2} + i\lambda$, that $L_1$ is antihermitian. 

\subsubsection{Scale Transformation (ST)}

Consider the scale transformation given by, 

\begin{equation}
D(\alpha) z = z' = e^{\alpha} z \ , \ 
\Lambda =
\begin{pmatrix}
a & b \\
c & d 
\end{pmatrix}
= 
\begin{pmatrix}
e^{\frac{\alpha}{2}} & 0 \\
0 & e^{-\frac{\alpha}{2}}
\end{pmatrix}
\end{equation}

It is easy to check that,

\begin{equation}
U \big(D(\alpha)\big) = e^{i \alpha K_2} = e^{\alpha L_0}
\end{equation}

So, 

\begin{equation}
{L_0}^{\dagger} = - L_0
\end{equation}

So in differential operator form, 

\begin{equation}
L_0 = i K_2 = - z\frac{d}{dz} - \Delta^*
\end{equation}

%Again one can check that $L_0$ is antihermitian. 

\subsubsection{Value of the Casimir for $\Delta$-Representation}

The Casimir of the Lorentz group is given by,

\begin{equation}
C = - (J_0)^2 + (K_1)^2 + (K_2)^2
\end{equation}

Let us now consider a representation with a specific value of $\Delta$. According to our previous calculation, in such a representation, 

\begin{equation}
J_0 = \frac{i}{2} (z^2 + 1) \frac{d}{dz} + i \Delta^* z 
\end{equation}

\begin{equation}
K_1= \frac{i}{2} (z^2 - 1) \frac{d}{dz} + i \Delta^* z 
\end{equation}

\begin{equation}
K_2 = i z \frac{d}{dz} + i \Delta^*
\end{equation}

%All of these operators are Hermitian with respect to the inner product given in Eq-$\ref{subinner}$. 

Now one can easily check that, 

\begin{equation}
C \Psi(\Delta,z) = \bigg( - (J_0)^2 + (K_1)^2 + (K_2)^2 \bigg) \Psi(\Delta,z) = \Delta\Delta^* \ \Psi(\Delta,z) = \big(\lambda^2 + \frac{1}{4}\big) \Psi(\Delta,z)
\end{equation}

So, 

\begin{equation}
\boxed{C = \Delta^* \Delta = \frac{1}{4} + \lambda^2} \ ,  \ \Delta = \frac{1}{2} + i\lambda , \ \lambda\in\mathbb{R}
\end{equation}

This is the value of the Casimir obtained in \cite{bargman}.

\subsubsection{Comments on The Representation of The Conformal Algebra}

The algebra of Lorentz generators in $(2+1)$ dimensions can be written as,

\begin{equation}
\big[L_0 , L_1\big] = - L_1 , \ \big[L_0 , L_{-1}\big] = L_{-1} , \ \big[L_1, L_{-1}\big] = 2 L_{0}
\end{equation}

This is also the algebra of global conformal group $SL(2,\mathbb{R})$ in one dimension. But as we have seen the reality properties of the generators are different if we think of the $SL(2,\mathbb{R})$ as acting unitarily on the Hilbert space of massless single particle states of a $(2+1)$ dimensional QFT. For example in the standard highest weight representation of the conformal group,

\begin{equation}
L_0^{\dagger} = L_0 , \ L_1^{\dagger} = L_{-1}, \  L_{-1}^{\dagger} = L_1
\end{equation}

whereas in our case unitarity requires that,

\begin{equation}\label{newtarity}
L_0^{\dagger} = - L_0 , \ L_1^{\dagger} = - L_1 , \ L_{-1}^{\dagger} = - L_{-1} 
\end{equation} \\

%\newpage
\section{ (3+1) Dimensions}

%\textit{In this section \textbf{bar} denotes complex conjugation unless otherwise specified}. \\

In four dimensions the same argument goes through unchanged except that the states now acquire an extra \textit{helicity} index. For the sake of convenience we will collect the necessary formulas. 

In four dimensions we can associate a hermitian matrix with a four momentum $P^{\mu} = (P^0,P^1,P^2,P^3)$ as, 

\begin{equation}
P = 
\begin{pmatrix}
P^0 - P^3 & P^1 + i P^2 \\
P^1 - i P^2 & P^0 + P^3 
\end{pmatrix} , \ \det P = -P^2
\end{equation}

In $(3+1)$ dimensions $SL(2,\mathbb{C})$ is the double cover of the Lorentz group $SO(3,1)$ and an $SL(2,\mathbb{C})$ matrix $\Lambda$ acts on $P$ as,

\begin{equation}
P \rightarrow P' = \Lambda P \Lambda^{\dagger}, \ 
\Lambda =
\begin{pmatrix}
a & b \\
c & d
\end{pmatrix} , \
ad-bc =1
\end{equation}

In four dimensions the space of null \textit{directions} is a two-sphere (a space-like cross section of the future light-cone in the momentum space). The stereographic coordinate of the two-sphere can be defined as, 

\begin{equation}
z = \frac{P^1 + i P^2}{P^0 + P^3}
\end{equation}

Here the projection is from the south-pole of the sphere so that the north-pole has coordinate $z=0$, which corresponds to the family of null vectors, $\big\{(E,0,0,E) \big| E>0 \big\}$, pointing in the positive $3$-direction. One can also check that under Lorentz transformation $\Lambda$, $z$ transforms as,

\begin{equation}
z \rightarrow z' = \frac{{P'}^1 + i {P'}^2}{{P'}^0 + {P'}^3} = \frac{az + b}{cz + d}
\end{equation}

Now if we introduce spherical polar coordinates in momentum space then we can write a null vector $P^{\mu}$ as, 

\begin{equation}
P^{\mu} = (E, E \sin\theta \cos\phi, E \sin\theta \sin\phi, E \cos\theta)
\end{equation}

In this prametrization $(\theta,\phi)$ become coordinates on the two-sphere. Its relation to the stereographic coordinates $z$ is given by, 

\begin{equation}
z = \tan\frac{\theta}{2} e^{i\phi}
\end{equation}

\subsection{Little Group of a Null Momentum Direction}

In four dimensions the little group of the standard null direction $\bigg\{(E,0,0,E) , E>0 \big| z=0\bigg\}$ is generated by the elements $\big( J_3, K_3, A, B\big)$ where $A = J_2 - K_1$ and $B = -J_1 - K_2$.  The commutators are given by, 
\begin{equation}
[A,B] = 0 \ ,  [J_3, A] = iB, \ [J_3, B] = - iA, \ [J_3, K_3] =0 , \ [K_3, A] = iA , \ [K_3, B] = iB 
\end{equation}

\subsection{Construction Of New Basis}

Let us define the following state, 

\begin{equation}\label{SST}
\ket{\lambda, \sigma, z= 0,\bar z= 0} = \frac{1}{\sqrt{8\pi^4}} \int_{0}^{\infty} dE \ E^{i\lambda} \ket{E,0,0,E \ ; \sigma}, \ \lambda\in\mathbb{R}
\end{equation} 
where $\sigma$ is the helicity. The momentum states have the standard normalisation given by, 
\begin{equation}
\bra{p_1,\sigma_1}\ket{p_2,\sigma_2} = (2\pi)^3 2|\vec p_1| \delta^3(\vec p_1 - \vec p_2) \delta_{\sigma_1 \sigma_2}
\end{equation}

The action of the little group on the state $\ket{\lambda, \sigma, z=0, \bar z=0}$ is given by,

\begin{equation}
A \ket{\lambda, \sigma, z= 0,\bar z= 0} = B \ket{\lambda, \sigma, z= 0,\bar z= 0} =0
\end{equation}

\begin{equation}
U\big(R_3(\phi)\big) \ket{\lambda, \sigma, z= 0,\bar z= 0} = e^{i\sigma\phi} \ket{\lambda, \sigma, z= 0,\bar z= 0}
\end{equation}

\begin{equation}
U\big(B_3(\eta)\big) \ket{\lambda, \sigma, z= 0,\bar z= 0} = e^{\eta\Delta} \ket{\lambda, \sigma, z= 0,\bar z= 0}, \  \Delta = 1+ i\lambda
\end{equation} 
where $\phi$ is the angle of rotation around the 3-axis and $\eta$ is the rapidity of the boost in the 3-direction. \\

Now let us define the states $\ket{\lambda,\sigma,z,\bar z}$ as,
\begin{equation}
\boxed{
\ket{\lambda,\sigma,z , \bar z} = \bigg(\frac{1}{1+ z\bar z}\bigg)^{1+i\lambda} \ U\big(R(z,\bar z)\big) \ket{\lambda,\sigma,z=0,\bar z=0} = N(z,\bar z) \ U\big(R(z,\bar z)\big) \ket{\lambda,\sigma,z=0,\bar z=0}}
\end{equation}

where $U\big(R(z,\bar z)\big)$ is a unitary rotation operator defined as, 

\begin{equation}
U\big(R(z, \bar z)\big) =  e^{-i\phi J_3} \ e^{-i\theta J_2} \ e^{i\phi J_3}, \ z = \tan\frac{\theta}{2} e^{i\phi}
\end{equation}

The rotation operator we have defined takes the standard null direction $\big\{ (E,0,0,E) \big | z=0\big\}$ to the direction given by $\bigg\{ (E,E\sin\theta\cos\phi,E\sin\theta\sin\phi,E\cos\theta) \bigg | z= \tan\frac{\theta}{2} e^{i\phi}\bigg\}$. \\

Now with our choice of normalization the inner product between the states is given by, 

\begin{equation}
\bra{\lambda_1,\sigma_1,z_1, \bar z_1}\ket{\lambda_2,\sigma_2,z_2,\bar z_2} = \delta(\lambda_1-\lambda_2) \delta^2(z_1 - z_2) \delta_{\sigma_1\sigma_2}
\end{equation}

where $\delta^2(z_1 - z_2) = \delta(\Re z_1 - \Re z_2) \delta(\Im z_1 - \Im z_2)$. 
 
\subsection{Action of The Lorentz Group}

Now we want to compute $U(\Lambda)\ket{\lambda,\sigma, z, \bar z}$. The computation is straightforward and we will summarize only the essential elements with obvious notation. Just as before,

\begin{equation}
\begin{aligned}
U(\Lambda) \ket{\lambda,\sigma, z, \bar z} 
& = N(z,\bar z) U(\Lambda) U(R(z,\bar z))\ket{\lambda,\sigma, z =0, \bar z =0} \\
& = N(z,\bar z) U(\Lambda R(z,\bar z))\ket{\lambda,\sigma, z =0, \bar z=0} \\
&= N(z,\bar z) U(R(\Lambda z,\Lambda \bar z)) U(R^{-1}(\Lambda z, \Lambda \bar z) \Lambda R(z,\bar z))\ket{\lambda,\sigma, z =0, \bar z=0} \\
& = N(z,\bar z) U(R(\Lambda z,\Lambda \bar z)) U(W(\Lambda, z, \bar z))\ket{\lambda,\sigma, z =0, \bar z=0}
\end{aligned}
\end{equation}

Now the Lorentz transformation $W(\Lambda, z, \bar z)$ belongs to the little group of the standard null direction $\big\{(E,0,0,E)\big | z=0\big\}$ and we need to calculate this. 

We will evaluate this by multiplying the corresponding $SL(2,\mathbb{C})$ matrices. Please see the appendix for our convention. 

The matrices are given by, 

\begin{equation}
R(z,\bar z) = 
\begin{pmatrix}
\cos\frac{\theta}{2} & e^{i\phi} \sin\frac{\theta}{2} \\
- e^{-i\phi} \sin\frac{\theta}{2} & \cos\frac{\theta}{2}
\end{pmatrix}
= \frac{1}{\sqrt{(1+|z|^2)}}
\begin{pmatrix}
1 & z \\
-\bar z & 1
\end{pmatrix} , \
z = \tan\frac{\theta}{2} e^{i\phi}
\end{equation}

\begin{equation}
\Lambda = 
\begin{pmatrix}
a & b \\
c & d
\end{pmatrix}
\in SL(2,\mathbb{C})
\end{equation}

\begin{equation}
R^{-1}(\Lambda z,\Lambda\bar z)  
= \frac{1}{\sqrt{(1+|\Lambda z|^2)}}
\begin{pmatrix}
1 & -\Lambda z \\
\Lambda\bar z & 1
\end{pmatrix} , \
\end{equation}

where,

\begin{equation}
\Lambda z = \frac{az + b}{cz + d} , \ \Lambda \bar z = \frac{\bar a \bar z + \bar b}{\bar c \bar z + \bar d}
\end{equation} 

If we multiply these matrices the little group element $W(\Lambda, z, \bar z)$ can be written as, 

\begin{equation}
W (\Lambda, z, \bar z) = 
\begin{pmatrix}
e^{\frac{\alpha}{2}}  & 0 \\
0 & e^{-\frac{\alpha}{2}}
\end{pmatrix}
\begin{pmatrix}
1 & 0 \\
-\beta & 1
\end{pmatrix}
= D(\alpha) S(\beta)
\end{equation}

where 

\begin{equation}
e^{\alpha} = \frac{1+ |z|^2}{1+ |\Lambda z|^2} \frac{1}{(cz+d)^2} 
\end{equation}

We do not need the expression of $\beta$. So we get, 

\begin{equation}
U(W) = U(D(\alpha)) U(S(\beta))
\end{equation}

Now it is easy to check that ,

\begin{equation}
U(D(\alpha)) = e^{\alpha L_0 + \bar\alpha \bar L_0}, \ D(\alpha) z = e^{\alpha} z
\end{equation}

where 

\begin{equation}
L_0 = \frac{i K_3 - J_3}{2} , \  \bar L_0 = - {L_0}^{\dagger}
\end{equation}

Similarly, 

\begin{equation}
U(S(\beta)) = e^{\beta L_1 + \bar\beta \bar L_1}, \ S(\beta) z = \frac{z}{1- \beta z}
\end{equation}

where, 

\begin{equation}
L_1 = \frac{i K_1 - J_1}{2} + i \frac{i K_2 - J_2}{2}, \ \bar L_1 = - {L_1}^{\dagger}
\end{equation}

For later use let us also write down the generator of translation in $z$. It is given by, 

\begin{equation}
U(T(\gamma)) = e^{\gamma L_{-1} + \bar\gamma \bar L_{-1}}, \  T(\gamma) z = z + \gamma
\end{equation}

where 

\begin{equation}
L_{-1} = \frac{J_1 - i K_1}{2} - i \frac{J_2 - i K_2}{2} , \  \bar L_{-1} = - {L_{-1}}^{\dagger}
\end{equation}

Now let us take note of the fact that the generator of the special conformal transformations on $z$ can be written as, 

\begin{equation}
L_1 = \frac{1}{2} (B - iA), \ \bar L_1 = - \frac{1}{2} (B + i A)
\end{equation}

where $A$ and $B$ are the elements of the little group of the null vector $(E,0,0,E)$. These are the generators which are set to zero on the states $\ket{E,0,0,E ; \sigma}$ to satisfy the requirement of a finite number of polarization states of a massless particle. Using these facts we get, 

\begin{equation}
\begin{aligned}
U(\Lambda)\ket{\lambda, \sigma, z, \bar z} 
&= N(z,\bar z) U(R(\Lambda z, \Lambda \bar z)) U(W(\Lambda, z , \bar z)) \ket{\lambda, \sigma, 0,0} \\
& = N(z,\bar z) U(R(\Lambda z, \Lambda \bar z)) U(D(\alpha)) U(S(\beta)) \ket{\lambda, \sigma, 0,0}  \\
& = N(z,\bar z) U(R(\Lambda z, \Lambda \bar z)) U(D(\alpha)) \ket{\lambda, \sigma, 0,0} \\
& = N(z,\bar z) U(R(\Lambda z, \Lambda \bar z)) (e^{\alpha})^{L_0} (e^{\bar\alpha})^{\bar L_0}  \ket{\lambda, \sigma, 0,0} \\
& = N(z,\bar z) U(R(\Lambda z, \Lambda \bar z)) \bigg[\frac{1+ |z|^2}{1+ |\Lambda z|^2} \frac{1}{(cz+d)^2}\bigg]^{\frac{\Delta - \sigma}{2}} \bigg[\frac{1+ |z|^2}{1+ |\Lambda z|^2} \frac{1}{(\bar c \bar z+\bar d)^2}\bigg]^{\frac{\Delta + \sigma}{2}} \ket{\lambda, \sigma, 0,0} \\
& = \frac{1}{(1 + |z|^2)^{\Delta}} \frac{(1 + |z|^2)^{\Delta}}{(1 + |\Lambda z|^2)^{\Delta}} \frac{1}{(cz+d)^{2h}} \frac{1}{(\bar c \bar z + \bar d)^{2 \bar h}} U(R (\Lambda z, \Lambda \bar z)) \ket{\lambda, \sigma,0,0} \\
& = \frac{1}{(cz+d)^{2h}} \frac{1}{(\bar c \bar z + \bar d)^{2 \bar h}} \ket {\lambda, \sigma, \Lambda z, \Lambda \bar z}
\end{aligned}
\end{equation}

where we have used the definition of the states $\ket{\lambda,\sigma,z,\bar z}$ and have defined,
\begin{equation}
\boxed{h = \frac{\Delta - \sigma}{2} = \frac{1 + i\lambda - \sigma}{2}, \ \bar h = \frac{\Delta + \sigma}{2} = \frac{1 + i\lambda + \sigma}{2}}
\end{equation}
Here $\bar h$ is \textit{not} the complex conjugate of $h$. 

Therefore, if we rename the state $\ket{\lambda,\sigma,z,\bar z}$ as $\ket{h,\bar h, z, \bar z}$, we can write, 

\begin{equation}
\boxed{U(\Lambda) \ket{h, \bar h, z, \bar z } = \frac{1}{(cz+d)^{2h}} \frac{1}{(\bar c \bar z + \bar d)^{2 \bar h}} \ket{h,\bar h, \Lambda z, \Lambda\bar z} = \bigg(\frac{d\Lambda z}{dz}\bigg)^{h}\bigg(\frac{d\Lambda \bar z}{d\bar z}\bigg)^{\bar h} \ket{h,\bar h, \Lambda z, \Lambda \bar z} ,} 
\end{equation}

where $\Lambda\in SL(2,\mathbb{C})$. Using this transformation law one can easily check that, 

\begin{equation}
\bigg( U(\Lambda) \ket{h',\bar h', z', \bar z'}, U(\Lambda) \ket{h,\bar h, z, \bar z} \bigg) = \bigg(\ket{h',\bar h', z', \bar z'}, \ket{h,\bar h, z, \bar z} \bigg)
\end{equation}

So the basic kets form a unitary representation of the Lorentz group and transform \textit{like (Lorentz) $SL(2,\mathbb{C})$ primary operators of weight $(h,\bar h)$ on the plane}. 

\subsection{Reality Condition on The Generators}

The Lorentz group acts on the two-sphere as the group of global conformal transformation. In the last section we decomposed the six generators as $\big(L_{-1}, L_0, L_{1}\big)$ and $\big(\bar L_{-1}, \bar L_0, \bar L_{1}\big)$ following the standard convention of 2-D CFT. Then we saw that the unitarity imposes the reality condition, 

\begin{equation}
{L_n}^{\dagger} = - \bar L_n , \  {\bar L_n}^{\dagger} = - L_n, \  n= -1, 0, 1
\end{equation}

The commutator of these generators are given by, 

\begin{equation}
[L_m, L_n] = (m-n) L_{m+n} , \  [\bar L_m, \bar L_n] = (m-n) \bar L_{m+n}, \ [L_m, \bar L_n] = 0
\end{equation}

So in this case unitarity imposes a reality condition on the conformal generators which is different from the reality condition in standard highest-weight representation. 
%Now suppose the global conformal algebra gets enhanced to the full virasoro algebra given by, 

%\begin{equation}
%[L_m,L_n] = (m-n) L_{m+n} + \frac{c_1}{12} m(m^2 - 1) \delta_{m+n}, \ m,n \in\mathbb{Z}
%\end{equation}

%\begin{equation}
%[\bar L_m,\bar L_n] = (m-n) \bar L_{m+n} + \frac{c_2}{12} m(m^2 - 1) \delta_{m+n} , m,n \in\mathbb{Z}
%\end{equation}

%\begin{equation}
%[L_m , \bar L_n] = 0
%\end{equation}

%where $(c_1 , c_2)$ are the central charges. Now it is natural to guess that we need a unitary representation of the Virasoro algebra subject to the following reality conditions, 

%\begin{equation}
%{L_n}^{\dagger} = - \bar L_n , \ {\bar L_n}^{\dagger} = - L_n , \  n \in\mathbb{Z}
%\end{equation}

%This reality condition requires that, 

%\begin{equation}
%{c_1}^{\dagger} = - c_2
%\end{equation}

%So the central charge $c_2$ is related to the complex conjugate of the central charge $c_2$. So we can write Virasoro algebras as, 

%\begin{equation}
%[L_m,L_n] = (m-n) L_{m+n} + \frac{c}{12} m(m^2 - 1) \delta_{m+n}, \ m,n \in\mathbb{Z}
%\end{equation}

%and 

%\begin{equation}
%[\bar L_m,\bar L_n] = (m-n) \bar L_{m+n} - \frac{c^{*}}{12} m(m^2 - 1) \delta_{m+n} , m,n \in\mathbb{Z}
%\end{equation}

%where $c^{*}$ is the complex conjugate of $c$.

%\begin{equation}
%(k^{\mu}, u=0) \overset{R(u)}\rightarrow (R(u) k^{\mu}, u'=R(u) u = \tan\frac{\theta}{2}) \overset{\Lambda R(u)}\rightarrow (\Lambda R(u) k^{\mu} , u'' = \Lambda R(u) u = \tan\frac{\Lambda\theta}{2})
%\end{equation}

\section{Action of The Space-time Translation Operators}

Let us now discuss the action of the space-time translations on the states $\ket{h,\bar h, z, \bar z}$. This will also give us the physical/geometrical interpretation of these states. 

There are four space-time translation operators given by, $P^{\mu} = \big( P^0 = H, P^1, P^2, P^3 \big)$. Here $H$ is the Hamiltonian - the generator of time translation. The generators transform as four-vector under Lorentz transformation, i.e, 

\begin{equation}
U(\Lambda)^{-1} P^{\mu} U(\Lambda) = {\Lambda^{\mu}}_{\nu} P^{\nu}
\end{equation}

Let us consider the following family of time dependent (Heisenberg-Picture) states defined as, 

\begin{equation}
\ket{h,\bar h, u, z, \bar z} = e^{iHu} \ket{h, \bar h, z, \bar z} = \bigg(\frac{1}{1 + z \bar z}\bigg)^{1+i\lambda} \ e^{iHu} \ U(R(z, \bar z))\ket{h, \bar h, 0, 0}
\end{equation}

where $u$ is a real number. These states are somewhat analogous to the Heisenberg picture states $\ket{t,\vec x}$ in the non-relativistic quantum mechanics. But now we also have the "internal degrees of freedom" parametrized by $(h,\bar h)$ or $(\lambda,\sigma)$. The completeness relation can be rewritten as, 
\be
\int_{-\infty}^{\infty} d\lambda \int d^{2}z \ket{h,\bar h, u, z, \bar z}\bra{h, \bar h, u, z, \bar z} =1
\ee

This is analogous to the completeness relation $\int d^n \vec x \ket{t,\vec x}\bra{t,\vec x} =1$ in non-relativistic quantum mechanics. We can calculate the matrix elements of the unitary translation operators in the basis $\big\{\ket{h, \bar h, z, \bar z}\big\}$ using its definition given in terms of the standard Wigner states. But for physical interpretation it will be more useful to have the transformation law of the time dependent states $\{\ket{h,\bar h,u,z,\bar z}\}$ under Poincare transformations. In other words we first construct the three parameter family of states and then see how they look like from a different frame of reference. Let us first state the results. 

Under a Lorentz transformation $\Lambda$, 

\begin {equation}\label{PP1}
\boxed{U(\Lambda) \ket{h, \bar h, u, z, \bar z} = \frac{1}{(cz+d)^{2h}} \frac{1}{(\bar c \bar z + \bar d)^{2 \bar h}} \ket{h \ , \bar h \ , \frac{u \ (1+ z \bar z)}{|az+b|^2 + |cz+d|^2}, \frac{az+b}{cz+d} \ , \frac{\bar a \bar z + \bar b}{\bar c \bar z + \bar d}}}
\end{equation}

and under space-time translation through a vector $l^{\mu}$, we get

\begin{equation}\label{PP2}
\boxed{e^{-il.P} \ket{h, \bar h, u, z, \bar z} = \ket{h, \bar h, u+ f(z, \bar z, l), z, \bar z}}
\end{equation}

where 

\begin{equation}\label{TR}
f(z, \bar z, l) = \frac{(l^0 - l^3) - (l^1 - i l^2) z - (l^1 + i l^2) \bar z + (l^0 + l^3) z \bar z}{1 + z \bar z}
\end{equation}
From Eq-$\ref{PP1}$ and Eq-$\ref{PP2}$ we can see that on the Heisenberg-Picture states $\ket{h,\bar h,u,z,\bar z}$ Poincare action is realized completely \textit{geometrically} modulo the little group factors. In fact we can think of a space parametrized by three coordinates $(u,z,\bar z)$ on which the Poincare group action is given by, 
\begin{equation}
\Lambda (u,z, \bar z) =  \bigg( \frac{u \ (1+ z \bar z)}{|az+b|^2 + |cz+d|^2} \ , \frac{az+b}{cz+d} \ , \frac{\bar a \bar z + \bar b}{\bar c \bar z + \bar d}\bigg)
\end{equation}
\begin{equation}
T(l) (u, z, \bar z) = (u + f(z, \bar z, l), z, \bar z)
\end{equation}

where $f$ is given above in Eq-$\ref{TR}$. \textit{This action of the Poincare group on the $(u,z,\bar z)$ space is the same as the action of the Poincare group at null-infinity in Minkowski space if we identify $(u, z, \bar z)$ with the Bondi coordinates}. We would like to emphasize that \textit{the coordinates $(u,z,\bar z)$ arise purely group-theoretically} and the coordinate $u$ has a time-like character because we get it by unitary Hamiltonian evolution. Also there is no Poincare invariant metric in the $(u,z,\bar z)$ space. 

Therefore the Heisenberg picture states $\ket{h,\bar h,u,z,\bar z}$ transform like the \textit{primary of the (Asymptotic) Poincare or inhomogeneous $SL(2,\mathbb{C})$ denoted by $ISL(2,\mathbb{C})$}.

From a physical point of view, in the basis $\ket{h,\bar h,z,\bar z}$ for single-particle quantum states, massless particles can be thought of as \textit{living at null-infinity}. More importantly, the \textit{Poincare invariant dynamics takes place at infinity}. This has the flavour of "holography". We will come back to this later. 

\subsection{Derivation}

We will not give the details of the algebra - leading to Eq-$\ref{PP1}$ and Eq-$\ref{PP2}$ - which is identical to whatever we have done before in the case Lorentz transformations. Let us just mention that the state $\ket{h,\bar h, z=0 , \bar z=0}$ has the following crucial property which is required for this thing to work. It satisfies, 
\begin{equation}
e^{ia(H - P^3)} \ket{h,\bar h, z=0 , \bar z=0} = e^{iaP^1} \ket{h,\bar h, z=0 , \bar z=0} = e^{iaP^2} \ket{h,\bar h, z=0 , \bar z=0} = \ket{h,\bar h, z=0 , \bar z=0}
\end{equation}

where $a$ is a real number and , 
\begin{equation}
\ket{h, \bar h, z= 0,\bar z= 0} = \frac{1}{\sqrt{8\pi^4}} \int_{0}^{\infty} dE \ E^{i\lambda} \ket{E,0,0,E \ ; \sigma}
\end{equation}

This is the same \textit{standard state} given in Eq-$\ref{SST}$ defined at the beginning except that we have replaced $(\lambda, \sigma)$ with $(h, \bar h)$. 

\subsubsection{A Geometric Argument}\label{G}

Let us now give a geometric argument which makes it clear that why we obtained this result, i.e, why we get an action of the Poincare group on a three dimensional space instead of the bulk four dimensional Minkowski space-time. This is related to the nature of the basis states and the geometry of the (conformally compactified) Minkowski space-time. 

So let us consider the state $\ket{h,\bar h, z=0 , \bar z=0}$ but now from the point of view of the Poincare group. We know that the  little group of a null direction in the momentum space is generated by four elements, $\big\{J_3, K_3, A = J_2 - K_1, B = -J_1 - K_2\big\}$ and as has already been discussed the state $\ket{h,\bar h, z=0 , \bar z=0}$ can be associated with the null direction given by, $\big\{ (E,0,0,E) , E>0 \big| z=0, \bar z=0\big\}$. Now when space-time translations are included one can check from the definition that the three translations $\big\{e^{ia(H-P^3)}, e^{iaP^1}, e^{iaP^2}\big\}$ keep the state $\ket{h,\bar h, z=0 , \bar z=0}$ \textit{invariant}. So let us adjoin the three generators to the little (algebra) group of a null direction. As a result we get a subgroup of the Poincare group generated by seven elements - $\big\{ J_3, K_3, A = J_2 - K_1, B = - J_1 - K_2, H-P^3, P^1, P^2\big\}$. The seven generators are closed under commutation. Let us denote the subgroup generated by them by $NH$.

\textit{The geometric significance of $NH$ is that it is isomorphic to the little group of a null-hyperplane in Minkowski space}. It is easy to check this.

% whose null-geodesic generators are pointing in the null direction given by $\big\{(E,0,0,E), E>0 \big\}$. Here we are assuming some embedding of the null-cone of the momentum space in the four dimensional Minkowski space written, for example, in retarded Bondi coordinates. 

For example let us consider the null-hyperplane passing through the origin of the Minkowski space. The equation of this null-hyperplane is, 

\begin{equation}
X.P = 0 , \ P^2 =0
\end{equation}

Here $P$ is a representative from the equivalence class $\big\{ P \sim \alpha P \big| \alpha > 0\big\}$ specifying the direction of the null normal.  The null geodesic generators of the null hyperplane are also parallel to $P$.  

Now let us consider a Lorentz transformation $\Lambda$, in the active sense, which maps the point $X$ lying on the null hyperplane to $\Lambda X$. Now if $\Lambda X$ also belongs to the null hyperplane then, $(\Lambda X) . P = 0 = X. (\Lambda^{-1} P)$. Therefore we must have, 

\begin{equation}
\Lambda^{-1} P \sim P \rightarrow \Lambda^{-1}P = \alpha \ P 
\end{equation}
for some $\alpha>0$. If we impose $\alpha = 1$ and choose $P$ to be $(1,0,0,1)$ then we get the standard little group of massless particles generated by $\big\{ J_3, A, B\big\}$. But since any other positive value of $\alpha$ is allowed, boost in the 3-direction is also an element of the little group. In this way we get four Lorentz generators, $\big\{ J_3, K_3, A, B\big\}$ as elements of the little group of a null-hypersurface. 

Now for translations given by, $X \rightarrow X' = X + a$, the condition that $X'$ belongs to the null-hyperplane gives
\begin{equation}
a.P = 0
\end{equation} 

This has three solutions corresponding to translations in the direction of $P$ itself and in the two space-like directions orthogonal to $P$. These are generated by $\big\{ H-P^3 , P^1, P^2\big\}$ if we take $P$ to be $(1,0,0,1)$. 

So altogether we get seven generators $\big\{ J_3, K_3, A, B, H-P^3, P^1, P^2\big\}$ which \textit{map the null-hyperplane to itself} and we can associate the state $\ket{h,\bar h, z=0 , \bar z=0}$ with the null-hypersurface $X.P=0$ with $P\sim(1,0,0,1)$. 

The remaining three generators of the Poincare group act non-trivially on the space of null-hyperplanes. Now a null-hyperplane can be thought of as the past light-cone of a point of the future null-infinity (or the future light-cone of a point of the past null-infinity)\footnote{See for example \cite{Penrose:1967wn}.}. So the space of null-hyperplanes in Minkowski space can either be thought of as the future null-infinity or the past null-infinity. The number of parameters also match. There are three-parameter family of null hyperplanes corresponding to the three Poincare generators which act non-trivially on the space of null hyperplanes and the null infinity is also three dimensional. This shows that we can think of the massless particle, described by a quantum state associated with a null-hyperplane, \textit{as sitting at one point at null infinity}. 

Let us now analyse the state $\ket{h,\bar h, u, z, \bar z} = e^{iHu} U(R(z,\bar z))\ket{h,\bar h,z=0,\bar z=0}$. As we have already described, we can associate the state $\ket{h,\bar h,z=0,\bar z=0}$ with the null-hyperplane $X.P=0$ with $P\sim (1,0,0,1)$ which passes through the origin. Now $U(R(z,\bar z))$ rotates the state and the rotated state corresponds to the null-hyperplane $X.P'=0$ where $P'= R(z,\bar z) P$. In this way we generate the 2-parameter family of null-hyperplanes all of which pass through the origin. They represent the family of states $\big\{\ket{h,\bar h,z,\bar z}\big\}$ with varying $(z,\bar z)$. Now the rest of the null hyperplanes are generated by time translating this two parameter family of null hyperplanes. This is essentially the action of $e^{iHu}$ on the states $\ket{h,\bar h,z,\bar z}$. This explains that why the states $\ket{h,\bar h, u, z,\bar z}$ transform under the asymptotic Poinacre group.

In passing we would like to point out that the generators $\big\{ J_3, K_3, A = J_2 - K_1, B = - J_1 - K_2, H-P^3, P^1, P^2\big\}$ of the Poincare group are also the \textit{kinematic generators in the Light-front quantization}. It will be interesting to see if the states in the light-front quantization are related to the asymptotic states $\ket{h,\bar h,u,z,\bar z}$ by some non-local transformation.

\section{Creation And Annihilation Fields At Null-Infinity}\label{ON}

In this section we change our notation a little bit and define,

\begin{equation}
\ket{\lambda,\sigma, z, \bar z} = \ket{h,\bar h, z, \bar z}, \ h = \frac{1+ i\lambda - \sigma}{2} , \  \bar h = \frac{1+ i\lambda + \sigma}{2}
\end{equation}

Let us now introduce Heisenberg-Picture creation operator  $A^{\dagger}_{\lambda,\sigma}(u,z,\bar z)$ corresponding to the states $\ket{\lambda,\sigma, u, z, \bar z}$ such that, 

\begin{equation}\label{AL}
%\begin{center}
%\begin{aligned}
U(\Lambda) \ A^{\dagger}_{\lambda,\sigma}(u, z,\bar z) \ U(\Lambda)^{-1} \\
 = \frac{1}{(cz+d)^{2h}} \frac{1}{(\bar c \bar z + \bar d)^{2 \bar h}}  \ A^{\dagger}_{\lambda,\sigma}\bigg(\frac{u \ (1+ z \bar z)}{|az+b|^2 + |cz+d|^2}, \frac{az+b}{cz+d} \ , \frac{\bar a \bar z + \bar b}{\bar c \bar z + \bar d}\bigg)
%\end{aligned}
%\end{center}
\end{equation}
and 
\begin{equation}\label{AT}
e^{-il.P} A^{\dagger}_{\lambda,\sigma} e^{il.P} = A^{\dagger}_{\lambda,\sigma} (u + f(z,\bar z,l), z, \bar z)
\end{equation}
where 
\begin{equation}
f(z, \bar z, l) = \frac{(l^0 - l^3) - (l^1 - i l^2) z - (l^1 + i l^2) \bar z + (l^0 + l^3) z \bar z}{1 + z \bar z}
\end{equation} 

Similarly the transformation property of the corresponding annihilation operator $A_{\lambda,\sigma}(u,z,\bar z)$ is given by,

\begin{equation}
%\begin{center}
%\begin{aligned}
U(\Lambda) \ A_{\lambda,\sigma}(u, z,\bar z) \ U(\Lambda)^{-1} \\
 = \frac{1}{(cz+d)^{2\bar h^{*}}} \frac{1}{(\bar c \bar z + \bar d)^{2 h^{*}}}  \ A_{\lambda,\sigma}\bigg(\frac{u \ (1+ z \bar z)}{|az+b|^2 + |cz+d|^2}, \frac{az+b}{cz+d} \ , \frac{\bar a \bar z + \bar b}{\bar c \bar z + \bar d}\bigg)
%\end{aligned}
%\end{center}
\end{equation}
and 
\begin{equation}
e^{-il.P} A_{\lambda,\sigma} e^{il.P} = A_{\lambda,\sigma} (u + f(z,\bar z,l), z, \bar z)
\end{equation}

where $*$ denotes complex conjugation.

Now the question is how are these creation/annihilation operators related to the standard creation/annihilation operators in the momentum eigenstate basis ? The simplest anwer is given by, 

\begin{equation}\label{amc1}
A^{\dagger}_{\lambda,\sigma} (u,z,\bar z) = \frac{1}{\sqrt{8\pi^4}} \bigg(\frac{1}{1+z \bar z}\bigg)^{1+i\lambda} \int_{0}^{\infty} dE \ E^{i\lambda} e^{iEu} \ a^{\dagger}(p,\sigma)
\end{equation}
\begin{equation}\label{amc2}
A_{\lambda,\sigma} (u,z,\bar z) = \frac{1}{\sqrt{8\pi^4}} \bigg(\frac{1}{1+z \bar z}\bigg)^{1-i\lambda} \int_{0}^{\infty} dE \ E^{-i\lambda} e^{-iEu} \ a(p,\sigma)
\end{equation}
where
\begin{equation}
p = E \bigg(1, \frac{z+\bar z}{1 + z\bar z},\frac{-i(z - \bar z)}{1+ z\bar z},\frac{1- z\bar z}{1+ z\bar z}\bigg), \  z = \tan\frac{\theta}{2} e^{i\phi}
\end{equation} 
and $a^{\dagger}(p,\sigma)$ is the creation operator in the momentum-helicity basis $\ket{p,\sigma}$.

This is essentially a rewriting of the relation between the basis states  
\begin{equation}
\ket{\lambda,\sigma, u, z, \bar z} = \frac{1}{\sqrt{8\pi^4}} \bigg(\frac{1}{1+z \bar z}\bigg)^{1+i\lambda} U\big(R(z,\bar z)\big) \int_{0}^{\infty} dE \ E^{i\lambda} e^{iEu} \ket{E,0,0,E,\sigma} 
\end{equation}
in second-quantized notation. 

The (anti) commutator between the creation and annihilation operators is given by, 

\begin{equation}\label{CA}
\begin{aligned} 
{[A_{\lambda,\sigma} (u, z, \bar z), A^{\dagger}_{\lambda',\sigma'}(u', z', \bar z')]_{\pm}} 
& = \bra{\lambda,\sigma, u, z, \bar z}\ket{\lambda',\sigma', u', z', \bar z'}  \\
&= \frac{\delta_{\sigma \sigma'}}{2\pi} \frac{\Gamma\big(i(\lambda' - \lambda)\big)}{(1 + z \bar z)^{i(\lambda' - \lambda)}} \frac{\delta^2(z' -z)}{\big(-i(u'-u + i0+)\big)^{i(\lambda' - \lambda)}}
\end{aligned}
\end{equation}

%\subsection{Reinterpretation}
%Now, given the transformation properties of the creation $A^{\dagger}_{\lambda,\sigma}(u, z, \bar z)$ or the annihilation operator $A_{\lambda,\sigma}(u, z, \bar z)$ we would like to interpret them as \textit{negative and positive frequency creation and annihilation fields living on null-infinity}. 

Eq-($\ref{amc1},\ref{amc2}$) and the transformation properties under Poincare group show that we should interpret $A_{\lambda,\sigma}(u,z,\bar z)$ and $A^{\dagger}_{\lambda,\sigma}(u,z,\bar z)$ as the \textit{positive and negative frequency annihilation and creation fields} living on \textit{null infinity} in the Minkowski space. So \textit{we have gone directly from the momentum space to the boundary of the Minkowski space-time}. In other words, $(z,\bar z)$ start as coordinates in the momentum-space but once we take into account the dynamics, $(z,\bar z)$ together with the time coordinate $u$ transmute into coordinates of the boundary of the Minkowski space-time. 

It will be interesting to relate this to the notion of asymptotic quantization \cite{Ashtekar:1978zz}. One difference is that here we do \textit{not} arrive at the quantum theory on the null-infinity by quantizing a \textit{classical theory living on null-infinity}. It will be interesting to clarify these things.

\section{Primary Of $ISL(2,\mathbb{C})$}

Let us consider a Poincare transformation $(l,\Lambda)$ which acts on Minkowski coordinates as $X^{\mu}\rightarrow {\Lambda^{\mu}}_{\nu} X^{\nu} + l^{\mu}$. 

Now for the sake of convenience we define a \textit{primary operator of (Poincare) $ISL(2,\mathbb{C})$} as any \textit{Heisenberg picture operator} $\phi_{h,\bar h}(u,z,\bar z)$ transforming as,
\be
U(\Lambda) \ \phi_{h,\bar h}(u, z,\bar z) \ U(\Lambda)^{-1} \\
 = \frac{1}{(cz+d)^{2h}} \frac{1}{(\bar c \bar z + \bar d)^{2 \bar h}}  \ \phi_{h,\bar h}\bigg(\frac{u \ (1+ z \bar z)}{|az+b|^2 + |cz+d|^2}, \frac{az+b}{cz+d} \ , \frac{\bar a \bar z + \bar b}{\bar c \bar z + \bar d}\bigg)
 \ee
\be
e^{-il.P} \phi_{h,\bar h}(u,z,\bar z) e^{il.P} = \phi_{h,\bar h} (u + f(z,\bar z,l), z, \bar z)
\ee

where 
\be\nonumber
f(z, \bar z, l) = \frac{(l^0 - l^3) - (l^1 - i l^2) z - (l^1 + i l^2) \bar z + (l^0 + l^3) z \bar z}{1 + z \bar z}
\ee

The $(h,\bar h)$ are defined to be, 
\be\nonumber
h = \frac{1+i\lambda - \sigma}{2}
\ee 
\be\nonumber
\bar h = \frac{1+ i\lambda + \sigma}{2}
\ee

We emphasize that the "primary" here does not refer to any highest-weight representation. The basic creation and annihilation fields at null-infinity that we have defined in the last section are examples of such primary operators. 

%From this point onward by \textit{primary} we will mean \textit{ $ISL(2,\mathbb{C})$ primary} unless otherwise specified.

\section{Constraints From Translational Invariance}\label{CT}
We will now work out some kinematical constraints on correlation functions of $ISL(2,\mathbb{C})$ primaries in Minkowski vacuum which follow from "bulk" translational invariance. Under space-time translation by an arbitrary four vector $l^{\mu}$, $u$ shifts by,

\begin{equation}\label{translation}
u \rightarrow u + f(z,\bar z,l) = u + \frac{(l^0 - l^3) - (l^1 - i l^2) z - (l^1 + i l^2) \bar z + (l^0 + l^3) z \bar z}{1 + z \bar z}\end{equation} 

whereas $z$ does not change. 

The correlation functions are invariant under (Poincare) $ISL(2,\mathbb{C})$ transformation, i.e, 

\begin{equation}
\bra{\Omega} \prod_{i=1}^n U(l,\Lambda) \ \phi_i(P_i) \ U(l,\Lambda)^{-1}\ket{\Omega} = \bra{\Omega} \prod_{i=1}^n \phi_i(P_i)\ket{\Omega}
\end{equation}

where $\phi_i (P_i)$ is some $ISL(2,\mathbb{C})$ primary operator inserted at the point $P_i=(u_i, z_i, \bar z_i)$ and $U(l,\Lambda)$ is a (Poincare) $ISL(2,\mathbb{C})$ transformation. $\ket{\Omega}$ is the Poincare invariant vacuum.

Let us start from the 4-point function.

\subsubsection{4-point function}

We denote the 4-point function by $A(P_1, P_2, P_3, P_4)$.  We first make a Lorentz transformation, $\Lambda$, to map the points $(z_1, z_2, z_3, z_4) \rightarrow (z, 1,0,\infty)$ where $z$ is the cross ratio of the four points given by, 

\begin{equation}
z = \frac{(z_1 - z_3)(z_2 - z_4)}{(z_1 - z_4)(z_2 - z_3)}
\end{equation}

Under this Lorentz transformation the $u$'s also transform to some other values say $u'$. So we can write, 

\begin{equation}
A (P_1, P_2, P_3, P_4) = G(\{\Lambda, z_i, \bar z_i, h_i, \bar h_i)\}) \ A (P_1', P_2',P_3',P_4')
\end{equation}
where 

\begin{equation}
P_1' = (u_1', z, \bar z) , \ P_2' = (u_2', 1,1) , \ P_3' = (u_3', 0,0), \ P_4' = (u_4' , \infty, \infty)
\end{equation}

and $G$ has \textit{no} $u$ dependence. Therefore we need to consider the 4-point function at these special values of $z_i$'s. Now translational invariance requires that, 

\begin{equation}
A(P_1', P_2',P_3',P_4') = A(P_1'', P_2'',P_3'',P_4'')
\end{equation}
where
\begin{equation}
\begin{aligned}
P_1'' = \big(u_1' + f(z, \bar z,l \big), z, \bar z), P_2'' = \big(u_2' + f(1,1,l), 1, 1\big), P_3'' = \big(u_3' + f(0,0,l), 0, 0\big), \\
P_4'' = \big(u_4' + f(\infty, \infty,l), \infty, \infty\big)
\end{aligned}
\end{equation}

Here we have used the fact that under translation $z_i$ does not change. For $z= 0, 1, \infty$ we have, 

\begin{equation}
f(0,0,l) = l^{0} - l^{3}, \ f(1,1,l) = l^{0} - l^{1} , \ f(\infty, \infty,l) = l^{0} + l^{3}
\end{equation}

Now if we take the translations to be infinitesimal then we get four differential equations given by, 

\begin{equation}
A_1 + A_2 + A_3 + A_4 = 0 
\end{equation}
\begin{equation}
A_1 - A_3 + \frac{1- z \bar z}{1 + z \bar z} A_4 = 0
\end{equation}
\begin{equation}
A_2 + \frac{z + \bar z}{1 + z \bar z} A_4 = 0
\end{equation}
\begin{equation}
\frac{z - \bar z}{1 + z \bar z} A_4 = 0
\end{equation}

where $A_i = \frac{\partial A}{\partial u_i'}$. 

Now if $z \ne \bar z$ then we get, $A_1 = A_2 = A_3 = A_4 = 0$. Therefore the most general non-trivil solution for the 4-point function can be written as, 

\begin{equation}
A(P_1, P_2, P_3, P_4) = \delta (\Im z) \ \tilde A(P_1, P_2, P_3, P_4)
\end{equation}

where $z$ is the the cross ratio of the four points $(z_1, z_2, z_3, z_4)$. So the four point function will be zero unless the cross ratio of the four insertion points are real. This is the same constraint obtained in \cite{Pasterski:2017ylz} from the study of Gluon scattering amplitude in the basis of conformal primary wave functions. Recently this has also been studied in detail in \cite{Lam:2017ofc}.

\subsubsection{3-point function}

In the case of the 3-point function one can repeat the same procedure except that now there is no cross ratio. We get 3 differential equations given by, 

\begin{equation}
A_1 + A_2 + A_3  = 0 
\end{equation}
\begin{equation}
A_1 - A_3  = 0
\end{equation}
\begin{equation}
A_2  = 0
\end{equation}

This has the trivial solution $A_1 = A_2 = A_3 = 0$. Therefore the most general non-trivial solution of the 3-point function is, 

\begin{equation} 
A(P_1, P_2, P_3) =  0
\end{equation}

This corresponds to the fact that in $(-+++)$ signature 3-point scattering amplitude of massless particles vanish for generic null momenta.

\subsubsection{2-point function}

The general solution of the 2-point function can be written as, 

\begin{equation}\label{2pt}
A(P_1, P_2) = N(\lambda_1,\lambda_2) \ \frac{\delta_{\sigma_1 +\sigma_2 , 0}}{(1+ z_1 \bar z_1)^{i(\lambda_1 + \lambda_2)}} \ \frac{\delta^2(z_1 - z_2)}{\big( u_1 - u_2)^{i(\lambda_1 + \lambda_2)}}
\end{equation}

where $N(\lambda_1,\lambda_2)$ is a prefactor which cannot be determined solely from the Poincare invariance. There should also be an $i\epsilon$ prescription to take care of the singularity at $u_1 = u_2$ but this cannot be determined just from symmetry consideration. 

%\pagebreak

\section{A Hint of Supertranslation }

Let us first consider the transition amplitude for a free massless particle \textit{in the $\ket{\lambda,\sigma,z,\bar z}$ basis}, given by 
\begin{equation}
\begin{aligned}
\bra{\lambda,\sigma,z,\bar z} e^{-iH(u-u')} \ket{\lambda',\sigma',z',\bar z'} 
&= \bra{\lambda,\sigma,u,z,\bar z}\ket{\lambda',\sigma',u',z',\bar z'} \\
&= \frac{\delta_{\sigma \sigma'}}{2\pi} \frac{\Gamma\big(i(\lambda' - \lambda)\big)}{(1 + z \bar z)^{i(\lambda' - \lambda)}} \frac{\delta^2(z' -z)}{\big(-i(u'-u + i0+)\big)^{i(\lambda' - \lambda)}}
\end{aligned}
\end{equation}

The Dirac delta function arises because a free particle does not change its direction \footnote{In this picture we can only talk about the direction of motion or the direction of the momentum vector because the magnitude has been integrated out.} of motion. Due to the presence of $\delta^{2}(z - z')$ the amplitude is manifestly invariant under space-time translations under which $u\rightarrow u + f(z,\bar z,l)$ and $z$ remains unchanged. Now it is easy to see that for the same reason \textit{the amplitude is in fact invariant under more general transformation} $u\rightarrow u+g(z,\bar z)$ where \textit{$g(z,\bar z)$ is an arbitrary smooth function on the sphere}. These can be \textit{identified} with \textit{BMS supertranslations} \cite{Bondi:1962px}. At this stage supertranslation invariance is just an \textit{accidental symmetry} which is manifest in this basis. 

The dependence of a general two point function on the coordinates of the null-infinity is completely fixed by the Poincare invariance. It is clear that the 2-point function given in Eq-$\ref{2pt}$ is also invariant under BMS supertranslations. This is also an accidental symmetry of the correlation function.

The supertranslation invariance of the transition amplitude of a free particle or the 2-point function is trivial from physical point of view \cite{Strominger:2013jfa}. In fact this cannot be the case for higher-point functions (in some interacting theory) because BMS is spontaneously broken in the Minkowski vacuum \cite{Strominger:2013jfa}. But, in some sense, this shows that perhaps it is "natural" to consider extension of the Poincare group to the BMS group once the field theory is formulated on null-infinity. This is also suggested by Strominger's conjecture about BMS invariance of the gravitational S-matrix \cite{Strominger:2013jfa} and recent advances  in understanding the relation between the infrared structure of gravity and asymptotic symmetry \cite{Strominger:2013lka,He,Kapec:2014opa, Strominger:2014pwa,Kapec:2016jld,Campiglia:2014yka,Campiglia:2015yka,Campiglia:2015kxa,Kapec:2017gsg,Strominger:2017zoo,Weinberg:1965nx,Cachazo:2014fwa,Schwab:2014xua,Bern:2014oka,Broedel:2014fsa,Avery:2015gxa,Sen:2017xjn,Sen:2017nim,Laddha:2017ygw,Chakrabarti:2017ltl,Chakrabarti:2017zmh,Laddha:2017vfh,Campoleoni:2017mbt,Hawking:2016msc} of flat space-time.

In fact the conjecture of \cite{Barnich:2009se,Banks:2003vp} and the recent works \cite{He,Kapec:2014opa, Strominger:2014pwa,Kapec:2016jld,Campiglia:2014yka,Campiglia:2015yka,Campiglia:2015kxa,Kapec:2017gsg,Bagchi:2016bcd,Bagchi:2016geg} suggest that that the superrotation symmetry plays a central role in a holographic reformulation of flat space physics. So the full symmetry group should perhaps be the \textit{(extended) BMS including superrotation}. The representation of the BMS group in three space-time dimensions has been studied in \cite{Barnich:2014kra,Campoleoni:2016vsh}.

%The dependence of the two point function on the coordinates of the null-infinity is completely fixed by the Poincare invariance. The resulting structure has some interesting properties which we want to point out. First of all, space-time translational invariance is manifest. Under translation $u\rightarrow u + f(z,\bar z)$ and due to the presence of $\delta^{2}(z_1 - z_2)$ the difference $(u_1 - u_2)$ remains invariant. Here $f(z,\bar z)$ is given by Eq-$\ref{translation}$.

%Secondly, for the same reason there is also a manifest infinite dimensional \textit{accidental} symmetry under which $u\rightarrow u+f(z,\bar z)$ where $f(z,\bar z)$ is now \textit{an arbitrary smooth function on the sphere}. These are essentially the \textit{BMS supertranslations}. 

We would like to emphasize that \textit{this BMS should not be thought of as an asymptotic symmetry group} \cite{Bondi:1962px, Banks:2003vp, Barnich:2009se}. There is no \textit{dynamical gravity} in this picture and in fact we did not require the presence of any "bulk space-time". The transition from momentum space to null-infinity was direct. Here the Supertranslation is some geometric transformation of the background space-time parametrized by $(u,z,\bar z)$ under which $2$-point function or the commutator (Eq-$\ref{CA}$) turns out to be symmetric. But so far \textit{the symmetry group which acts on states in the Hilbert space} is still the four dimensional Poincare group or $ISL(2,\mathbb{C})$. 

The fact that we get a hint of supertranslation in a \textit{purely non-gravitational setting} is perhaps an indication of \textit{holography in asymptotically flat space-time}. This is somewhat similar to the situation in AdS$_3$-CFT$_2$ correspondence. The infinite dimensional Virasoro symmetry of a two-dimensional conformal field theory can be understood in a purely field theory setting without dynamical gravity. But in asymptotically AdS$_3$ spaces Virasoro can also be understood as the asymptotic symmetry group of the bulk gravity theory. These two facts nicely fit together in AdS$_3$-holography. Similarly the goal here is to understand \textit{BMS in a purely non-gravitational setting}. If such an understanding can be reached in a consistent manner then that will perhaps be an indication of flat-space holography.

\section{Conformal Primary Wave-Functions With Translational Invariance}

For simplicity we consider the massless scalar conformal primary wave functions \cite{Pasterski:2017kqt, Cheung:2016iub, deBoer:2003vf}. Scalar conformal primary wave functions in (3+1) dimensions are given by,

\begin{equation}\label{PSS}
\Phi_{\Delta}^{\pm} (X^{\mu} | z, \bar z) = \bigg(\frac{1}{1 + z\bar z}\bigg)^{\Delta} \int_{0}^{\infty} dE  E^{\Delta -1} e^{\pm i E Q \cdot X} e^{-\epsilon E} = \bigg(\frac{1}{1 + z\bar z}\bigg)^{\Delta} \frac{(\mp i)^{\Delta} \Gamma(\Delta)}{(- Q \cdot X \mp i\epsilon)^{\Delta}} \ ,  \ \Delta = 1 + i\lambda
\end{equation}
where $X^{\mu}$ is an arbitrary point in the Minkowski space and $Q^{\mu}(z,\bar z)$ is a standard null vector given by
\begin{equation}
Q^{\mu}(z,\bar z) = \bigg(1, \frac{z + \bar z}{1 + z\bar z}, \frac{-i(z - \bar z)}{1 + z\bar z}, \frac{1- z\bar z}{1 + z\bar z}\bigg)
\end{equation}
which under a $SL(2,\mathbb{C})$ transformation transforms as ,
\begin{equation}
Q^{\mu}(\Lambda z, \Lambda \bar z) = \frac{1 + z\bar z}{|az + b|^2 + |cz + d|^2} \ {\Lambda^{\mu}}_{\nu} Q^{\nu} (z, \bar z), \ \Lambda z = \frac{az+b}{cz+d}, \ \Lambda \bar z = \frac{\bar a \bar z + \bar b}{\bar c \bar z + \bar d}
\end{equation}
where ${\Lambda^{\mu}}_{\nu}$ is the corresponding Lorentz transformation matrix. It was shown in \cite{Pasterski:2017kqt} that the wave-functions given in Eq-$\ref{PSS}$ form a complete set of delta function normalizable solutions of the massless Klein-Gordan equation if $\Delta = 1+ i\lambda$. 

Under Lorentz transformation the scalar conformal primary wave functions transform as,
\begin{equation}
\Phi_{\Delta}^{\pm} \bigg({\Lambda^{\mu}}_{\nu} X^{\nu} \bigg| \frac{az +b}{cz+d}, \frac{\bar a \bar z + \bar b}{\bar c \bar z + \bar d}\bigg) = (cz + d)^{\Delta} (\bar c\bar z + \bar d)^{\Delta} \ \Phi_{\Delta}^{\pm} (X^{\mu} | z, \bar z)
\end{equation} \\

To see the action of the space-time translation in a simple manner let us introduce an extra parameter $u$ and new wave functions labelled by $\Delta (= 1 + i\lambda)$ and three parameters $(u, z, \bar z)$. The new wave functions are given by, 
\begin{equation}\label{S}
\boxed{\Phi_{\Delta}^{\pm} (X^{\mu} | u, z, \bar z) = \bigg(\frac{1}{1 + z\bar z}\bigg)^{\Delta} \int_{0}^{\infty} dE  E^{\Delta -1} e^{\pm iE (Q \cdot X + u)} e^{-\epsilon E} = \bigg(\frac{1}{1 + z\bar z}\bigg)^{\Delta} \frac{(\mp i)^{\Delta} \Gamma(\Delta)}{(- Q \cdot X - u \mp i\epsilon)^{\Delta}}} \ 
\end{equation}
This still solves the massless Kelin-Gordon equation and for a \textit{fixed value of $u$} the family of wave functions parametrized by $(\Delta,u,z,\bar z)$ form a complete set just like the wave functions given in Eq-$\ref{PSS}$ . Now one can easily check that under Lorentz transformation, 
\begin{equation}
\Phi_{\Delta}^{\pm} \bigg({\Lambda^{\mu}}_{\nu} X^{\nu} \bigg| \frac{u (1 + z \bar z)}{ |az+b|^2 + |cz+d|^2}, \frac{az +b}{cz+d}, \frac{\bar a \bar z + \bar b}{\bar c \bar z + \bar d}\bigg) = (cz + d)^{\Delta} (\bar c\bar z + \bar d)^{\Delta} \ \Phi_{\Delta}^{\pm} (X^{\mu} | u, z, \bar z)
\end{equation}

and under spacetime translation by four vector $l^{\mu}$ ,

\begin{equation}
\Phi_{\Delta}^{\pm} (X^{\mu} + l^{\mu}  \big |  u + f(z, \bar z,l), z, \bar z) = \Phi_{\Delta}^{\pm} (X^{\mu} | u, z, \bar z)
\end{equation}
where 
\begin{equation}\label{pt}
f(z,\bar z,l) = \frac{(l^0 - l^3) - (l^1 - i l^2) z - (l^1 + i l^2) \bar z + (l^0 + l^3) z \bar z}{1 + z \bar z}
\end{equation}

So the wave functions $\Phi_{\Delta}^{\pm}(X^{\mu} | u,z,\bar z)$ transform in a simple manner under the Poincare group and it is natural to identify the parameters $(u,z,\bar z)$ as the Bondi coordinates of the null infinity in Minkowski space. 

The above procedure has a simple geometric interpretation. As we have described in section-($\ref{G}$) the states $\ket{h,\bar h, u,z, \bar z}$ are associated with null-hyperplanes in the Minkowski space. The original 2-parameter family of conformal primary wave functions, given by $\Phi_{\Delta}^{\pm} (X^{\mu} | z, \bar z)$ (Eq-$\ref{PSS}$), are singular along the null hyperplanes $X\cdot Q(z,\bar z) = 0$ passing through the origin. They correspond to the states $\ket{h,\bar h,u=0,z,\bar z}$. So to get the rest of the states with nonzero $u$ we have translate the family $X\cdot Q(z,\bar z)=0$ along the time axis. This gives rise to the wave functions $\Phi_{\Delta}^{\pm} (X^{\mu} |u, z, \bar z)$ (Eq-$\ref{S}$) singular along the null-hyperplanes $X\cdot Q(z,\bar z) + u =0$. This 3-parameter family exhausts all the null-hyperplanes in the Minkowski space and correspond to the family of states $\ket{h,\bar h,u,z,\bar z}$ located on null-infinity. This makes it clear that why the Poincare group has simple action on the wave-functions $\Phi_{\Delta}^{\pm} (X^{\mu} |u, z, \bar z)$.

Now suppose instead of plane-waves we use the wave functions $\big\{\Phi_{\Delta_i}^{\pm}(X^{\mu} | u_i,z_i,\bar z_i)\big\}$ as the wave functions of the external particles then we can write down a modified Mellin-transform as, 
\begin{equation}
\boxed{\tilde A(\{h_i, \bar h_i, u_i, z_i , \bar z_i\}) = \prod_{i=1}^{n} \bigg(\frac{1}{1 + z_i\bar z_i}\bigg)^{\Delta_i} \int_{0}^{\infty} dE_i \ E_{i}^{\Delta_i - 1} e^{-i\epsilon_i E_i u_i} A(\{E_i, z_i, \bar z_i\})}
\end{equation}

where $h_i = \bar h_i = \frac{\Delta_i}{2}$ and we have also defined $\epsilon_i = +1$ for an outgoing particle and $\epsilon_i=-1$ for an incoming particle. This is a slightly modified form of the Mellin amplitude defined in \cite{Pasterski:2016qvg,Pasterski:2017ylz}.

 $\tilde A$ amplitude behaves like \textit{correlation function of $ISL(2,\mathbb{C})$ primaries inserted at points at null infinity} and transform under the \textit{Asymptotic Poincare Group} as, 
\be
\tilde A\big(\{h_i, \bar h_i, \Lambda u_i, \Lambda z_i , \Lambda \bar z_i\}\big) = \prod_{i=1}^{n} (cz_i+d)^{2 h_i} (\bar c \bar z_i + \bar d)^{2\bar h_i} \ \tilde A\big(\{h_i, \bar h_i, u_i, z_i , \bar z_i\}\big)
\ee
\be
\tilde A\big(\{h_i, \bar h_i, u_i + f(z_i,\bar z_i,l), z_i , \bar z_i\}\big) = \tilde A\big(\{h_i, \bar h_i, u_i,  z_i , \bar z_i\}\big)
\ee
where 
\be{\nonumber}
\Lambda u_i = \frac{(1 + z_i\bar z_i) u_i}{|az_i+b|^2 + |cz_i+d|^2}
\ee

and $f(z,\bar z,l)$ is given by Eq-$\ref{pt}$. 

It will be interesting to compute the $\tilde A$ amplitude for some field theories along the lines of \cite{Pasterski:2017ylz,Cardona:2017keg,Banerjee:2017jeg,Schreiber:2017jsr}.

The $\tilde A$ amplitudes can be thought of as the "scattering amplitude with asymptotic states" given by $\ket{h_i,\bar h_i,u_i,z_i,\bar z_i}$. It is very likely that a relation of the type,
\be
\tilde A(\{h_i, \bar h_i, u_i, z_i , \bar z_i\}) \sim \bra{\Omega} \prod_{i=1}^n \phi_{(h_i,\bar h_i)}(u_i,z_i,\bar z_i)\ket{\Omega}
\ee
exists where $\phi_{(h_i,\bar h_i)}$ are $ISL(2,\mathbb{C})$ primaries at null-infinity and the correlator in the R.H.S is \textit{computed in a theory defined at null-infinity}. The correlation function should perhaps be \textit{u-ordered} because, as we have seen, the coordinate $u$ has time-like character. It will be interesting to make these things more precise.

\section{Acknowledgement}

It is a pleasure to thank  Arjun Bagchi, Nabamita Banerjee, Sayantani Bhattacharya, Bidisha Chakrabarty, Abhijit Gadde, Sudip Ghosh, Sachin Jain, Dileep Jatkar, Arnab Kundu, Alok Laddha, R.Loganayagam, Gautam Mandal, Sunil Mukhi, Amin Nizami, Partha Paul, Ashoke Sen, Yogesh Srivastava and Amitabh Virmani for useful discussions and Alok Laddha for useful comments on the draft. I would also like to thank the string theory groups in HRI Allahabad, ICTS Bengaluru, IISER Pune and TIFR Mumbai for hospitality at various stages of the work. I would like to thank the organizers and the participants of the National Strings Meeting 2017 held in NISER Bhubaneswar where part of this work was presented.

\end{document}